\documentclass[12pt]{article}
\usepackage{graphicx}
\usepackage{setspace}
\usepackage{epstopdf}
\usepackage{bm}
\usepackage{float}
\usepackage{amssymb}
\usepackage{amsmath}

\newcommand{\ket}[1]{ | #1 \rangle}
\newcommand{\bra}[1]{\langle #1 |}

\newcommand{\ua}{\uparrow}
\newcommand{\da}{\downarrow}

\begin{document}

\title{Controlling a quantum gas of polar molecules in an optical lattice}
\author{Jacob P. Covey, Steven A. Moses, Jun Ye,$^{\dagger}$ Deborah S. Jin$^{\ddagger}$\\
\\
\normalsize{JILA, National Institute of Standards and Technology and University of Colorado,} \\ \normalsize{and Department of Physics, University of Colorado, Boulder, CO 80309, USA}
\\
\normalsize{Corresponding authors: $^{\dagger}$Ye@jila.colorado.edu, $^{\ddagger}$Jin@jilau1.colorado.edu}}
\date{}

\maketitle

\tableofcontents

\begin{abstract}
The production of molecules from dual species atomic quantum gases has enabled experiments that employ molecules at nanoKelvin temperatures. As a result, every degree of freedom of these molecules is in a well-defined quantum state and exquisitely controlled. These ultracold molecules open a new world of precision quantum chemistry in which quantum statistics, quantum partial waves, and even many-body correlations can play important roles. Moreover, to investigate the strongly correlated physics of many interacting molecular dipoles, we can mitigate lossy chemical reactions by controlling the dimensionality of the system using optical lattices formed by interfering laser fields. In a full three-dimensional optical lattice, chemistry can be turned on or off by tuning the lattice depth, which allows us to configure an array of long-range interacting quantum systems with rich internal structure. Such a system represents an excellent platform for gaining fundamental insights to complex materials based on quantum simulations and also for quantum information processing in the future.
\end{abstract}

\section{Creation of ultracold molecules}

The field of ultracold polar molecules has exploded over the past decade as researchers have worked to extend the control offered in ultracold atom experiments producing atomic Bose-Einstein condensates (BECs)~\cite{Anderson1995, Davis1995, Bradley1995, Bradley1997} and Degenerate Fermi gases (DFGs)~\cite{DeMarco1999, Truscott2001, Dieckmann2002} to the world of molecules.  Heteronuclear molecules have electric dipole moments, and they provide strong, long-range, and anisotropic interactions with precise tunability.  A strong interest has emerged in the scientific community to study systems with long-range interactions. These systems are ideal candidates for the study of strongly correlated quantum phenomena, as well as for quantum simulation of lattice models relevant to some outstanding problems in condensed-matter physics.  The anisotropic nature of dipolar interactions provides powerful opportunities to control chemical reactions in the low energy regime, and dipolar interactions can also give rise to novel forms of quantum matter such as Wigner crystallization~\cite{Knap2012}, $d$-wave superfluidity in optical lattices~\cite{Kuns2011}, fractional Chern insulators~\cite{Yao2013}, and spin-orbit coupling~\cite{Syzranov2016}.

Efforts to produce stable gases of ultracold polar molecules started in the early 2000's, and are documented in a recent review~\cite{Carr2009}. However, the production of ground-state polar molecules in the quantum regime was a challenging task and it took a major effort to achieve the first success in 2008~\cite{Ni2008}. The key to the success was the combination of the use of a Feshbach resonance for magneto-association of bialkali atoms and coherent optical state transfer via stimulated Raman adiabatic passage (STIRAP).

Feshbach resonances have been studied extensively in the context of controlling atomic interactions via magnetic fields, and they have become one of the most powerful tools for quantum gas experiments. Homonuclear, weakly bound ``Feshbach" molecules were produced directly from single-species ultracold atomic samples by using such resonances~\cite{Donley2002, Jochim2003, Herbig2003}. In particular, homonuclear Feshbach molecules composed of fermionic atoms are themselves bosons, and their production enables studies of a crossover from the BCS-type Cooper pairing of fermions into a BEC of weakly bound molecules~\cite{Greiner2003, Zwierlein2003, Bertenstein2004}. In an optical lattice, such weakly-bound molecule formation gives rise to quantum Zeno suppression of collisions between molecules as they tunnel~\cite{Syassen2008}. Work on using the STIRAP technique to transfer molecules to more deeply bound states was performed on both Rb$_2$~\cite{Winkler2007,Lang2008} and Cs$_2$ molecules~\cite{Danzl2008,Danzl2009}.  However, homonuclear molecules do not have an electric dipole moment, and are thus not useful for studying dipolar scattering or long-range dipole-dipole interactions.

Inspired by the work with weakly bound homonuclear molecules, many researchers started producing heteronuclear dimers, involving various combinations of alkali atoms~\cite{Kohler2006,Chin2010}. However, the path to deeply bound, or ground-state, molecules was quite challenging. The earliest efforts for producing cold polar molecules relied on photoassociation to optically couple free atoms to an excited molecular electronic state, which then spontaneously decays to many rovibrational states in the ground electronic potential. In order to collect the population in the rovibrational ground state, one can choose an excited state with good Franck-Condon overlap~\cite{Franck1926, Condon1928} with both the free atom state and the ground rovibrational state~\cite{Jones2006}. However, the efficiencies for both the excitation from free atoms to an excited molecular state and the subsequent spontaneous decay to a specific rovibrational state in the ground are very low. As a consequence, the phase-space density of such cold molecule samples is typically below $\sim 10^{-10}$~\cite{Hudson2008}. Several more recent experiments have managed to improve on this limitation by using electronically excited states with stronger Franck-Condon coupling to deeply bound states (see e.g. References~\cite{Reinaudi2012, stellmer2012}), but they are not in the rovibrational ground state and the phase-space densities are still very far from reaching the quantum regime.

Alternatively, the coherent transfer approach first adiabatically converts a pair of free atoms of different species into a highly vibrationally excited state in the ground electronic potential using a Feshbach resonance, and then couples these Feshbach molecules to the rovibrational ground state via an adiabatic Raman transfer process through an intermediate electronic excited state. The first success came in 2008 with the production of heteronuclear fermionic KRb molecules~\cite{Ospelkaus2008,Ni2008}. Research on deeply bound heteronuclear molecules has exploded in recent years. Several groups have now produced heteronuclear molecules in their deeply bound ground state, in species such as bosonic RbCs~\cite{Takekoshi2014, Molony2014}, fermionic NaK~\cite{Park2015}, and bosonic NaRb~\cite{Guo2016}. The fermionic KRb molecules recently reached quantum degeneracy~\cite{Moses2015, Covey2016}.

\subsection{Magneto-association}
\begin{figure}
\centering
\includegraphics[width=3in]{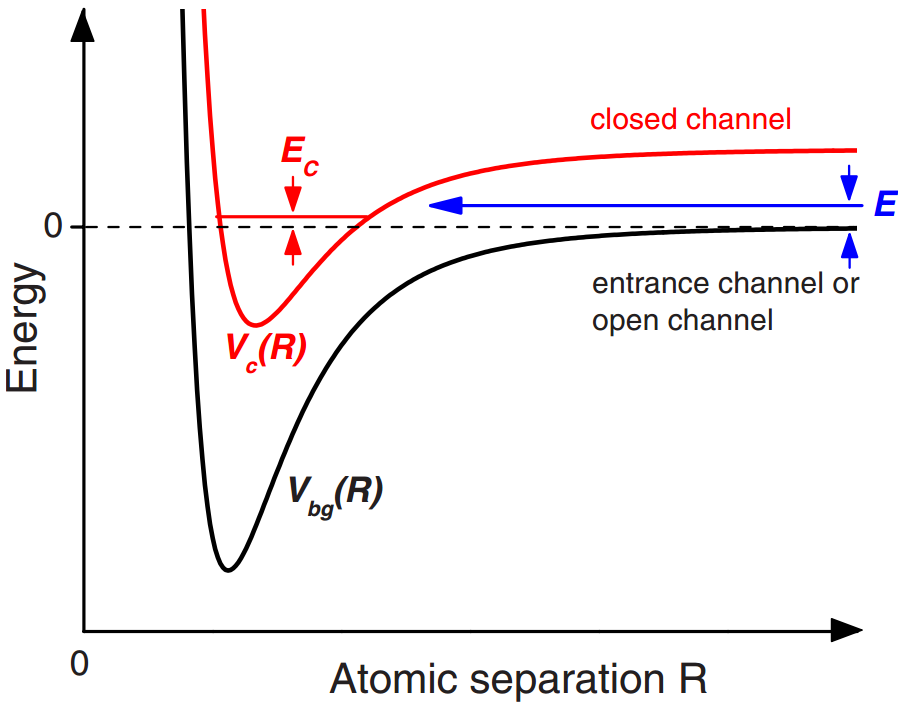}
\caption{Potential energy wells of the closed and open channel near a Feshbach resonance. The energy offset $E_\text{c}$ can be adjusted via a differential Zeeman shift using a magnetic field to realize resonant coupling of the two channels. Reproduced with permission from Reference~\cite{Chin2010}.}
\label{FBR1}
\end{figure}

A pair of free atoms can be directly converted to a molecule near the dissociation limit of the ground potential when the bound molecular state (closed channel) becomes degenerate in energy with the relative motion of the free atoms at large internuclear separations (open channel). This type of Feshbach resonance exists for essentially all atomic species with a magnetic moment, and the two-channel model for a Feshbach resonance is shown in Figure~\ref{FBR1}. Since the two channels typically have different magnetic moments, the energy difference between them, $E_\text{c}$, can be adjusted with a magnetic field to realize resonant coupling of the two channels~\cite{Chin2010}.

\begin{figure}
\centering
\includegraphics[width=3in]{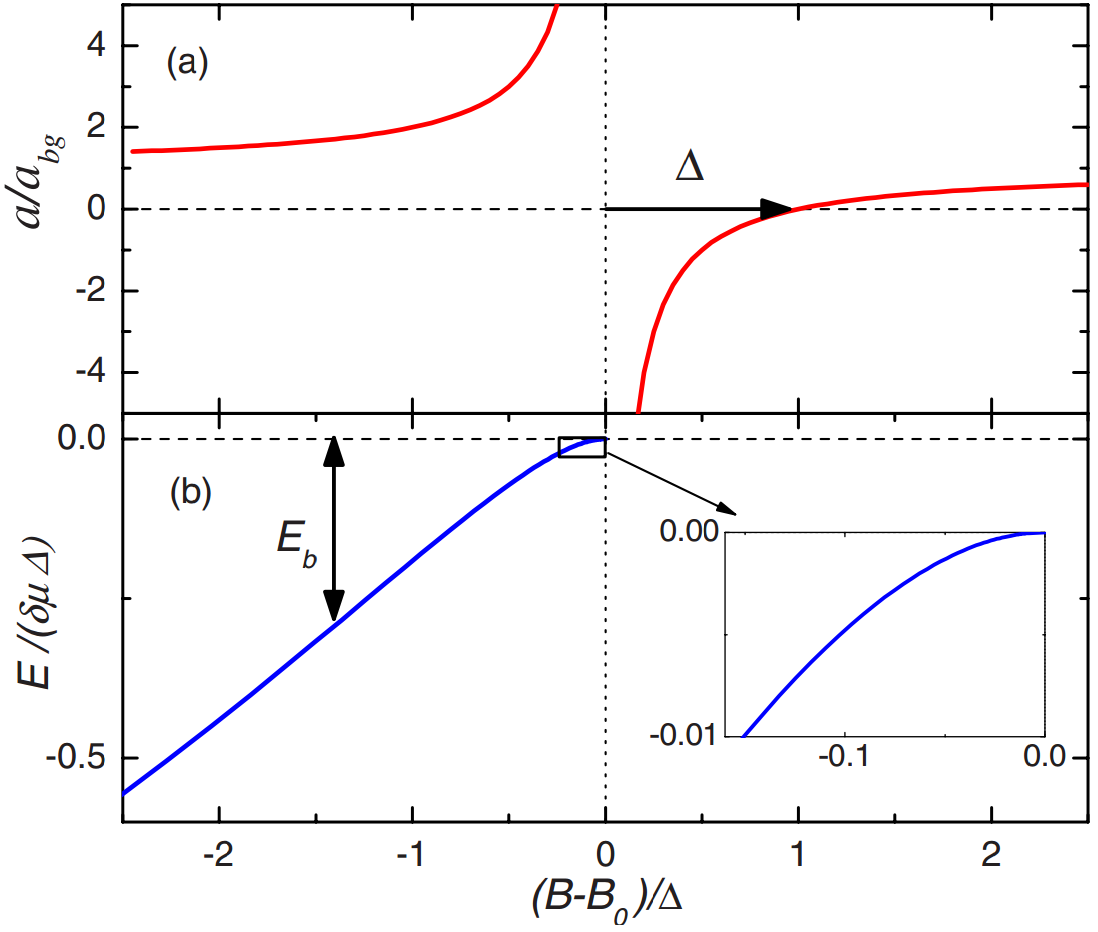}
\caption{Scattering length and energy of two atoms in the vicinity of a Feshbach resonance. (a) shows the scattering length normalized to the background scattering as a function of field, while (b) shows the binding energy as a function of field. Note that this figure assumes $a_\text{bg}>0$, which is not always the case. Reproduced with permission from Reference~\cite{Chin2010}.}
\label{FBR2}
\end{figure}

At a specific field $B_0$, the energy difference between the closed and open channel vanishes, and the resultant scattering length of the two atoms diverges. Tuning $B$ away from $B_0$ on the side of the positive scattering length leads to a molecular binding energy $E_\text{b}$, and it is given by
\begin{equation}
\label{eq:BindingEnergy}
E_\text{b}=\hbar^2/2 \mu a^2,
\end{equation}
where $\mu$ is the reduced mass of the atom pair and $a$ is the scattering length of the atoms~\cite{Chin2010}. Note that the energy scale plotted in Fig.~\ref{FBR2}b is normalized by $\delta \mu$, which is the difference in magnetic moments of the two channels. Typical binding energies in Feshbach molecule experiments are $\sim 0.1-1$ MHz. The scattering length as a function of field is shown in Fig.~\ref{FBR2}a, and is given by
\begin{equation}
\label{eq:ScatteringLength}
a(B)=a_\text{bg}\bigg(1-\frac{\Delta}{B-B_0}\bigg),
\end{equation}
where $a_\text{bg}$ is the background scattering length, and $\Delta$ is the width of the resonance, which is the difference between $B_0$ and the field that corresponds to $a=0$ (see Fig.~\ref{FBR2}a). Typical widths in the most experimentally used resonances are $\sim 1-300$ G, although some molecules (e.g. Cs$_2$~\cite{Herbig2003, Danzl2009}) use a resonance width of several mG.

\begin{figure}
\centering
\includegraphics[width=3in]{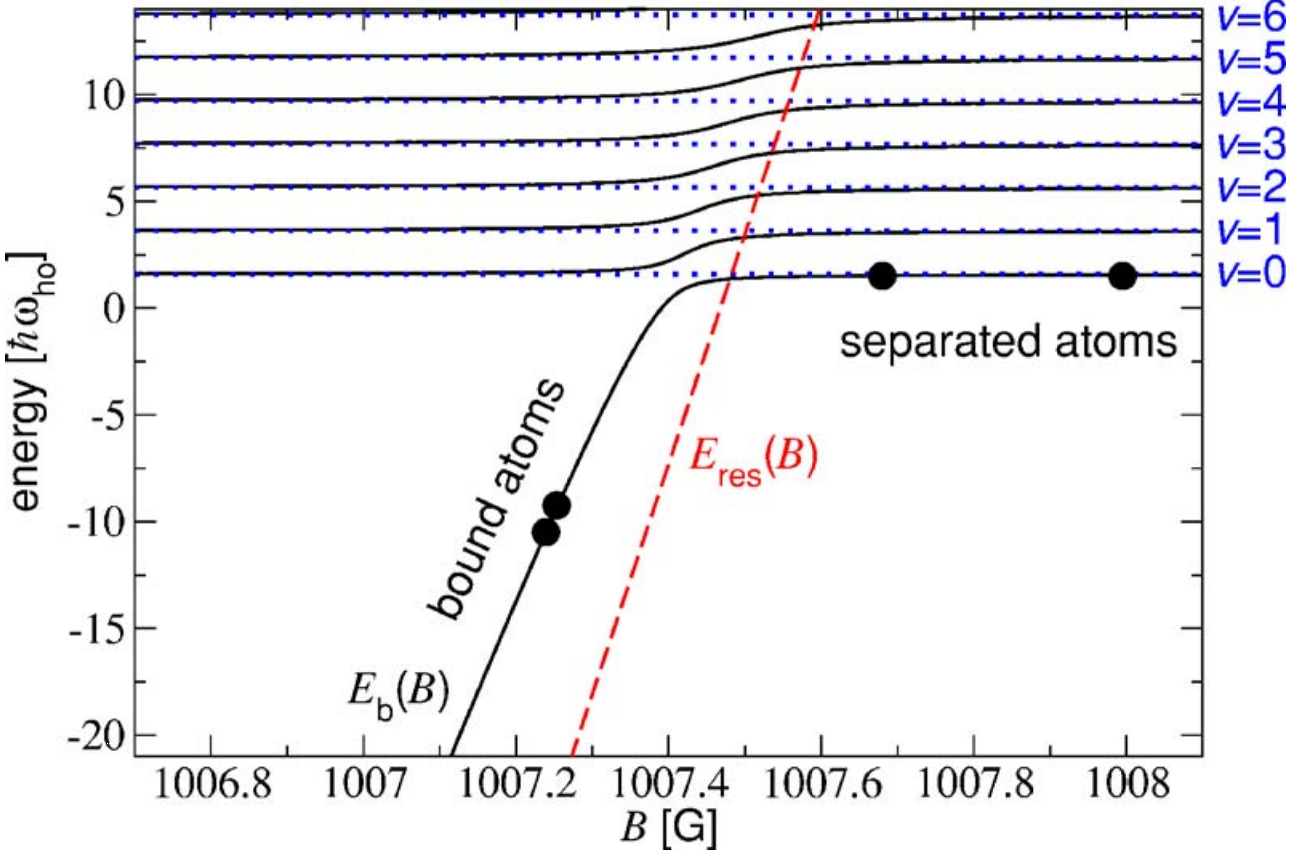}
\caption{The energy of two atoms as a function of magnetic field. The harmonic oscillator levels are shown above the resonance. The closed channel energy is $E_\text{res}(B)$. The lowest harmonic oscillator level couples to the Feshbach molecule, whose energy reduces below the resonance as $E_\text{b}(B)$.  Reproduced with permission from Reference~\cite{Kohler2006}.}
\label{figFBM}
\end{figure}

To create Feshbach molecules, one typically sweeps $B$ across the resonance from the attractive side ($a_\text{bg}<0$) to the repulsive side ($a_\text{bg}>0$) so that the pair of free atoms adiabatically enters the closed molecular channel. Figure~\ref{figFBM} illustrates this process schematically, and also shows the change of the molecular binding energy as a function of $B$. The free atoms are confined in a harmonic trap, and the different entrance levels correspond to different harmonic oscillator states in the trap. In a fully adiabatic process only the lowest energy state of the free atoms can couple to the bound channel, in the limit that the harmonic oscillator states are well resolved. This process can be well described by a Landau-Zener coupling mechanism, and the probability that free atoms are adiabatically converted to molecules is given by~\cite{Kohler2006}:
\begin{equation}
\label{eq:LZ1}
P_\text{FbM}=1- e^{-2\pi \delta_\text{LZ}}.
\end{equation}
Here $\delta_\text{LZ}$ is the Landau-Zener parameter, which in a weak harmonic trap is
\begin{equation}
\label{eq:LZ2}
\delta_\text{LZ}^\text{ho}=\frac{\sqrt{6} \hbar}{\pi 2 \mu a_\text{ho}^3} \left| \frac{a_\text{bg} \Delta}{\dot{B}} \right|,
\end{equation}
where $a_\text{ho}$ is the harmonic oscillator length of the molecule in the harmonic trap, and $\dot{B}$ is the sweep rate across the resonance.

\begin{figure}
\centering
\includegraphics[width=3in]{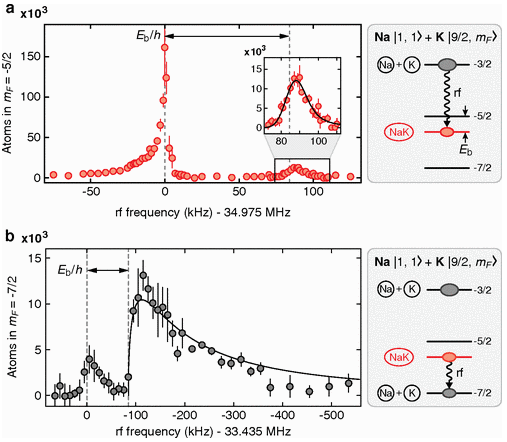}
\caption{RF association of NaK Feshbach molecules from two different initial states. In both cases the binding energy can be seen as a separate peak that can be spectroscopically resolved from the free K atom peak. Reproduced with permission from Reference~\cite{Wu2012}.}
\label{NaKFBM}
\end{figure}

An alternative method to create Feshbach molecules is to couple the open and closed channels with a radio frequency (rf) field when the two channels are tuned close to each other. This technique is used in many experiments, and it allows researchers to perform Rabi oscillations between free atoms and molecules~\cite{Donley2002,Olsen2009}. A beautiful illustration of this technique in NaK Feshbach molecule production is provided in Fig.~\ref{NaKFBM}, which is reproduced from Reference~\cite{Wu2012}.

\subsection{Coherent optical transfer}
After producing weakly bound Feshbach molecules, a two-photon-based coherent population transfer technique, named Stimulated Raman adiabatic passage (STIRAP)~\cite{Marte1991, Unanyan1998}, is used to transfer molecules to the rovibratonal ground state.  Three states are involved in the process: the initial Feshbach molecule state $\ket{f}$, an electronically excited state $\ket{e}$ (which is usually short-lived), and the rovibrational ground state $\ket{g}$.  The excited state is chosen, after an exhaustive spectroscopy search, to optimize the Franck-Condon factors with both $\ket{f}$ and $\ket{g}$.  The states are coupled with two laser fields: the up leg, with Rabi frequency $\Omega_u$, couples $\ket{f}$ and $\ket{e}$; and the down leg, with Rabi frequency $\Omega_d$, couples $\ket{g}$ and $\ket{e}$.  Although the Franck-Condon factors are reasonably large, the transition dipole moments are significantly smaller than typical cycling transitions in neutral atoms.  During the adiabatic and coherent process of STIRAP, a field-dressed dark state, $\cos \theta \ket{f} + \sin \theta \ket{g}$, where $\theta = \tan^{-1} \left( \frac{\Omega_u}{\Omega_d} \right)$, is strictly maintained. The laser intensities for the two beams are varied such that initially $\Omega_d/\Omega_u \gg 1$, and then the relationship adiabatically changes to $\Omega_u/\Omega_d \gg 1$. The laser intensities are ramped over a duration $\tau$ that is sufficiently long compared to the inverse of the Rabi frequencies. Since the population is trapped in the dark state, population transfer takes places from $\ket{f}$ to $\ket{g}$ without populating the lossy state $\ket{e}$. The typical STIRAP protocal for ultracold molecules is based on a dark resonance configuration, where both the one-photon and the two-photon detunings are zero. The excited state population is therefore adiabatically eliminated, as long as the following condition is satisfied: $\beta << 1/\tau << \Omega$, where $\beta$ is the relative laser linewidth of the two-photon transition~\cite{Yatsenko2014}. Thus, the two laser fields must be phase coherent with each other during the time $\tau$.  For typical pulse times $\sim 10 \, \mu$s, relative laser linewidths less than 1 kHz are required, which can be achieved by stabilizing the lasers to an optical frequency comb~\cite{Ospelkaus2008,Ni2008} or a high-finesse optical cavity~\cite{aikawa2009, gregory2015}.

\begin{figure}
\centering
\includegraphics[width=3in]{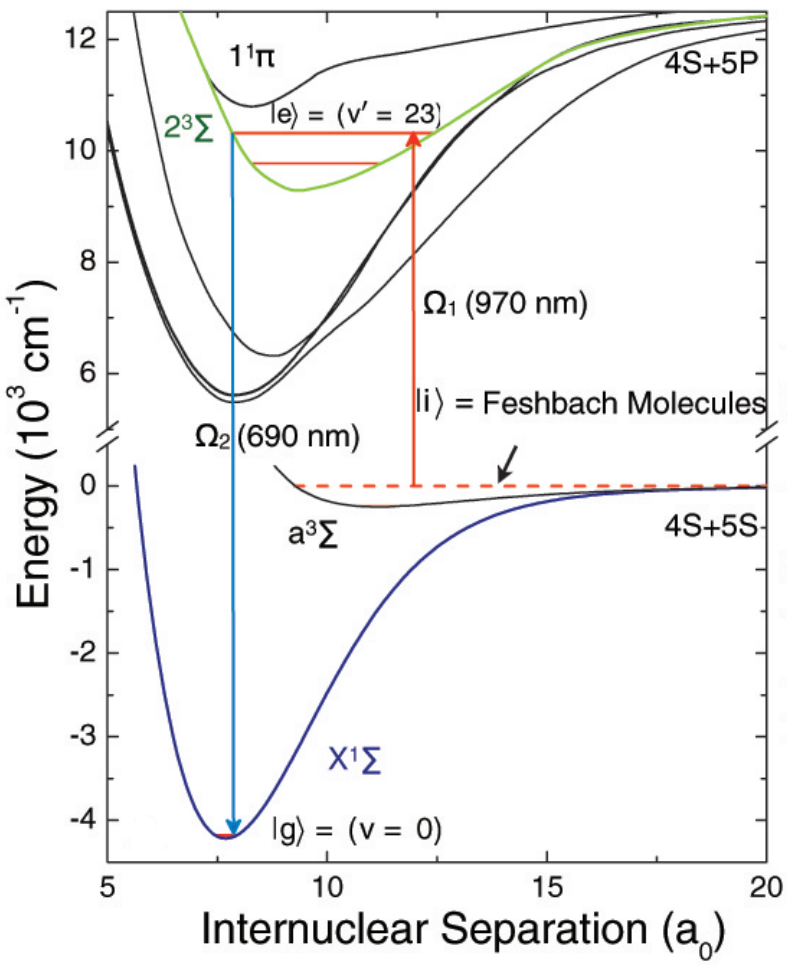}
\caption{Two-photon coherent state transfer for KRb from the weakly bound Feshbach molecule state $\ket{i}$ to the absolute molecular ground state $\ket{f}$ ($\nu$=0, $N = 0$ of X$^1\Sigma$). Reproduced with permission from Reference~\cite{Ni2008}.}
\label{KRbSTIRAP}
\end{figure}

Figure~\ref{KRbSTIRAP} shows the transfer states used for efficient production of ground-state KRb molecule. We note that the initial Feshbach state is the least bound vibrational level of the $^3\Sigma$ ground potential, while the rovibrational ground state is in the $^1\Sigma$ ground potential. Thus the intermediate electronic state must contain an admixture of singlet and triplet characters in order to allow reasonable transition strengths on both legs of the transfer. Such an admixture results from significant spin-orbital coupling in the excited electronic state. Using these states, we typically find a $\sim90\%$ one-way transfer efficiency to and from the ground state. An important aspect to consider in such a STIRAP sequence is that the difference in energy between the initial and final states is $\sim4000$ K, which is enormous compared to the temperature of the gas ($\sim100$ nK). Therefore, one might wonder whether any of this extra energy is deposited into the kinetic energy of molecules, which would heat the molecular gas out of the ultracold regime. However, since the transfer process is fully coherent, the energy difference between the states is carried away by the photons, and no kinetic energy (beyond an effective photon recoil via the co-propagating laser beams) is deposited into the molecular gas, which remains in the $\sim100$ nK regime. When producing molecules directly in an optical lattice (to be discussed later), this process allows the molecules to occupy the ground band of the lattice with very high probability~\cite{Danzl2010, Moses2015,Safavi-Naini2015}.

\section{Quantum-state controlled chemical reactions and dipolar collisions}
This section focuses primarily on ultracold KRb molecules, which is the only species for which detailed investigations of ultracold chemistry have been undertaken to date.  Immediately after ultracold polar KRb molecules were produced in the rovibrational ground state, it became clear that chemical reactions were leading to loss of molecules confined in a conventional far-off-resonance optical dipole trap~\cite{Ospelkaus2010b, Ni2010}.

\subsection{Control of the molecules' quantum state}

The STIRAP process transfers the molecules to the rovibrational ground state in the singlet ($X^1 \Sigma^{+}$) electronic ground potential. Although the electronic angular momentum is zero, there is still hyperfine structure arising from the nuclear magnetic moments of the constituent atoms.  K has nuclear quantum number $I^{\text{K}}=4$, while Rb has nuclear quantum number $I^{\text{Rb}}=3/2$, so there are 36 hyperfine states for each rotational state $\ket{N,m_N}$, where $N$ is the principal rotational quantum number and $m_N$ is its projection onto the quantization axis.  These states can be labelled as $\ket{N,m_N,m_\text{K},m_\text{Rb}}$, where $m_{\text{K}}$ and $m_{\text{Rb}}$ are the nuclear spin projection quantum number for K and Rb, respectively. In the ultracold regime, the initial Feshbach association creates weakly bound molecules in a single quantum state since K and Rb atoms approach each other in the lowest partial wave $s$ ($L$=0), where $\hbar L$ is the quantized relative angular momentum. Thus, through a combination of the two-photon angular momentum selection rule and the spectral resolution provided by the coherent STIRAP process, the ground-state molecules will also occupy a single hyperfine quantum state, which in the case of KRb is $\ket{0,0,-4,1/2}$. This state is not the lowest in energy in the hyperfine manifold. The absolute ground state is actually $\ket{0,0,-4,3/2}$.  To reach this absolute ground state and to also provide a general spectroscopy tool to coherently and efficiently transfer populations between different hyperfine states, we have developed another coherent two-photon transfer technique, which is based on the use of two microwave photons to couple from two different hyperfine states in $\ket{N = 0}$ to a common hyperfine state in the $\ket{N = 1}$ manifold via electric dipole transitions.  This two-photon coupling is enabled by the fact that the $\ket{N = 1}$ state has non-negligible (a few percent) spin mixing due to coupling between rotation and the nuclear electric quadrupole moment~\cite{Aldegunde2009, Ospelkaus2010a}.  This microwave-based Raman transition allows us to transfer the molecular population to any hyperfine state in $\ket{N = 0}$, including the absolute ground state $\ket{0,0,-4,3/2}$.  This technique is also important for the creation of a 50:50 mixture of two hyperfine states in the ground state or a 50:50 mixture of two rotational states for KRb molecules, which we use to study the role of quantum statistics of identical molecules in the reaction process.

\subsection{Role of quantum statistics on collisions}

Owing to the antisymmetrization requirement of the overall wavefunction for two identical fermions, spin-polarized (i.e., the same internal state) $^{40}$K$^{87}$Rb molecules collide via odd partial waves. At ultralow temperatures, the lowest odd partial wave is the $p$-wave ($L$=1). In contrast, distinguishable molecules (such as a hyperfine or a rotational mixture) can collide via $s$-wave ($L$=0). Thus, when molecules are brought to the quantum regime, their interactions and reactions will be governed by the molecular quantum statistics, single partial-wave scattering, and quantum threshold laws at a vanishing collision energy. The first set of experiments that explored this interesting new regime for molecular reactions is reported in Reference~\cite{Ospelkaus2010b}, which studied the dependence of reaction rate on temperature and quantum statistics without applying a laboratory electric field.

\begin{figure}
\begin{center}
\includegraphics[width=9cm]{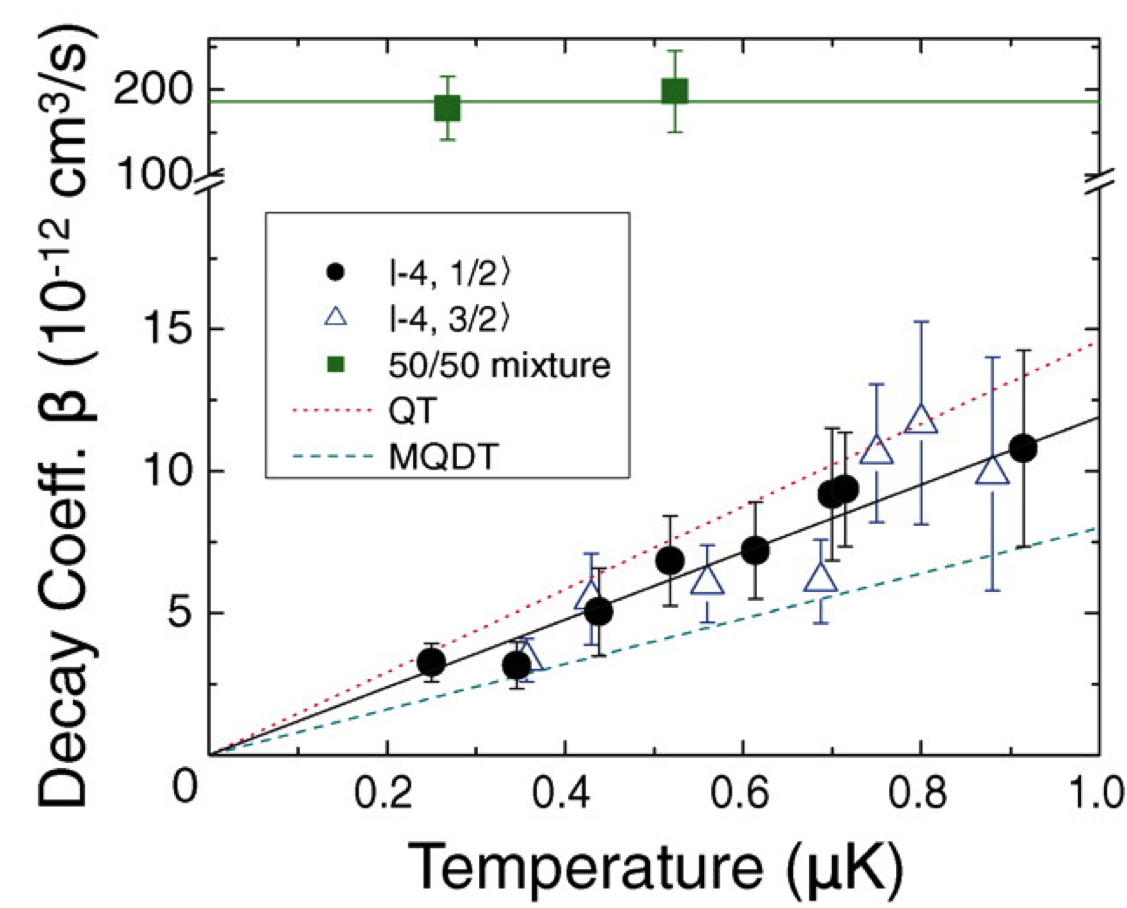}
\caption{Loss rate coefficient as a function of temperature in a bulk gas for spin-polarized KRb molecules, showing a linear increase with temperature characteristic of the $p$-wave threshold behavior. For comparison, loss rate for a 50:50 mixture of hyperfine spin states is shown, where the nature of $s$-wave collisions removes the temperature dependence for the rate constant while the loss magnitude is increased by a factor of 10-100. Reproduced with permission from Reference~\cite{Ospelkaus2010b}.}
\label{fig:tempdep}
\end{center}
\end{figure}

As a result of the chemical reactions, the number density $n$ of trapped molecules decreases according to
\begin{equation}
\frac{dn}{dt} = -\beta n^2 - \alpha n,
\end{equation}
where $\beta$ is the two-body loss coefficient and $\alpha$ describes a decrease in density arising from an increase in temperature. The temperature increase results from loss from the highest density part of the sample, where the particles have the lowest potential energy in a harmonic trap~\cite{Ospelkaus2010b}. The remaining one-body loss mechanism has a much longer time constant. The two-body loss rate $\beta$ is found to be proportional to $T$ for spin-polarized samples, which is the expected dependence from the Bethe-Wigner threshold laws for $p$-wave collisions (Fig.~\ref{fig:tempdep}). The lack of any oscillatory signatures in the rate coefficient dependence on temperature is consistent with a simple model that assumes that after the molecules tunnel through the $p$-wave barrier they chemically react with a near unity probability~\cite{Idziaszek2010}.

\begin{figure}
\begin{center}
\includegraphics[width=8cm]{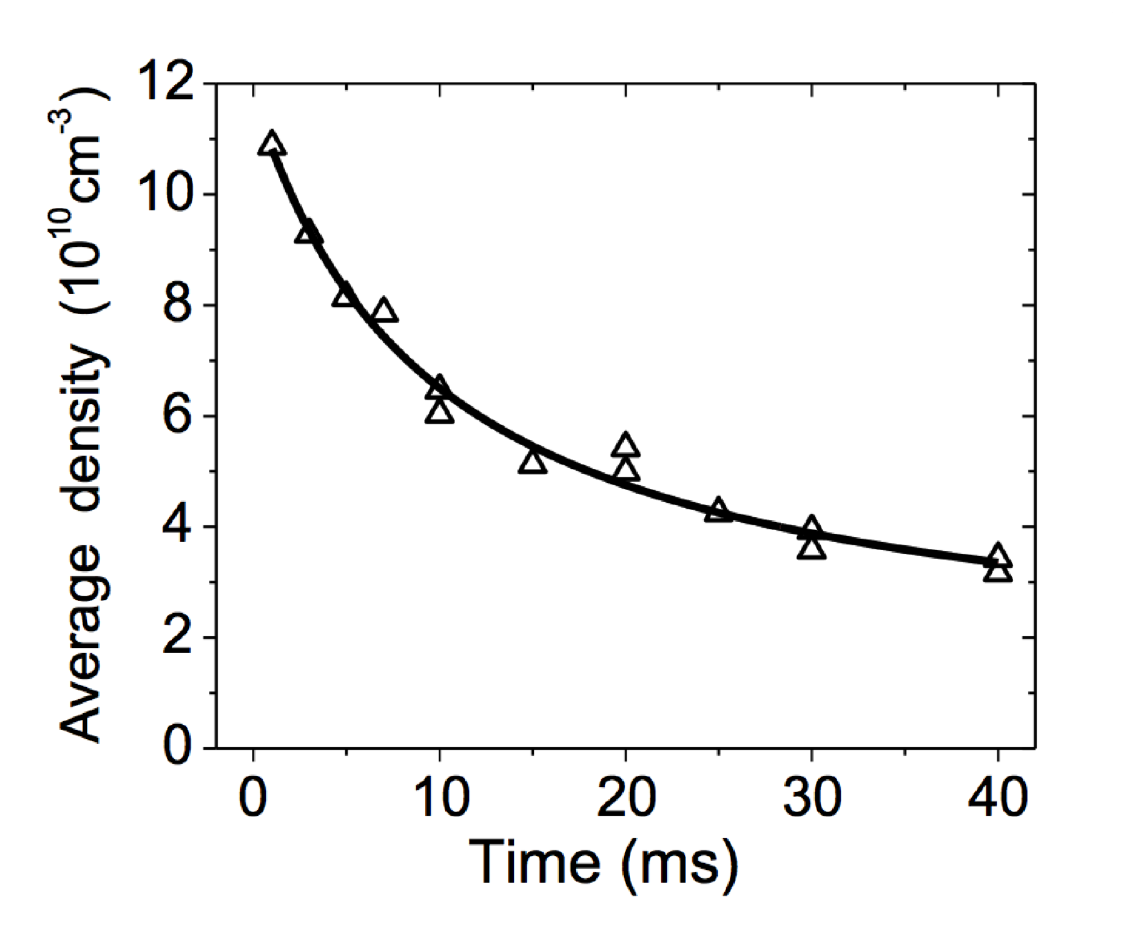}
\caption{Decay of an incoherent mixture of rotational states $\ket{0,0,-4,1/2}$ and $\ket{1,-1,-4,1/2}$, in a harmonic trap.  The extracted loss rate is $9.0(4) \times 10^{-10}$ cm$^3$ s$^{-1}$, which is almost 5 times higher than for the ground-state hyperfine mixture. Reproduced with permission from Reference~\cite{Yan2013}.}
\label{fig:mixtureloss}
\end{center}
\end{figure}

For an incoherent 50:50 mixture of two hyperfine states in the ground state ($\ket{0,0,-4,1/2}$ and $\ket{0,0,-4,3/2}$), $s$-wave collisions start to play a dominant role since the molecules from the two different states are distinguishable. As a result, the loss rate shows no measurable temperature dependence and its magnitude is increased by ten- to hundredfold in comparison to the $p$-wave rate. The measured rate agrees very well with the predicted universal loss rate related to the two-body van der Waals length.  We note that the absolute ground state ($\ket{0,0,-4,3/2}$) demonstrates the same loss rate as other higher lying hyperfine states. Molecules prepared in an incoherent 50:50 mixture of rotational states (half in $\ket{0,0,-4,1/2}$ and half in $\ket{1,-1,-4,1/2}$) have the largest loss rate observed at zero electric field.  A typical loss curve can be seen in Fig.~\ref{fig:mixtureloss}, from which a decay coefficient of $\beta= 9.0(4) \times 10^{-10}$ cm$^3$ s$^{-1}$ is extracted.  This is almost 5 times higher than for the $N=0$ mixture \cite{Yan2013}.  Finally, when both atoms and molecules are prepared in their lowest-energy states, we have also observed a universal $s$-wave loss rate from the exothermic atom-exchange reaction for K + KRb $\rightarrow$ K$_2$ + Rb. For an endothermic process of Rb + KRb $\rightarrow$ Rb$_2$ + K, which is forbidden, the observed loss rate approaches zero within the measurement uncertainty~\cite{Ospelkaus2010b}. However, if the Rb density becomes too high, a three-body loss process involving two Rb atoms and a KRb will limit the molecular lifetime.

\begin{figure}
\begin{center}
\includegraphics[width=7cm]{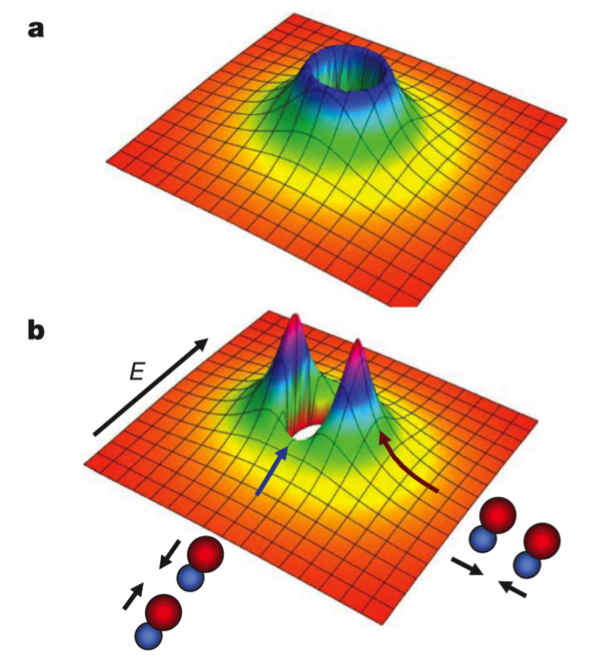}
\caption{(a) The $p$-wave barrier at zero electric field is isotropic and does not depend on the orientation in which the molecules collide.  (b) In an applied DC electric field, dipolar interactions reduce the centrifugal barrier for ``head-to-tail" collisions (blue arrow) and increase the centrifugal barrier for ``side-by-side" collisions.  Reproduced with permission from Reference~\cite{Ni2010}.}
\label{fig:pwave}
\end{center}
\end{figure}

When an electric field is applied, the molecules develop an electric dipole moment in the laboratory frame and this modifies how the molecules interact with each other. First, the isotropic centrifugal barrier arising from the $p$-wave becomes anisotropic. For $L=1$, the projection onto the electric field-defined quantization axis, $m_L$, dictates how the molecules collide: $m_L=0$ corresponds to a ``head-to-tail" collision, which is attractive under the dipole-dipole interaction; while $m_L= \pm 1$ corresponds to a ``side-by-side" collision, which is repulsive under the dipole-dipole interaction. This is schematically shown in Fig.~\ref{fig:pwave}. Furthermore, the dipole-dipole interaction also mixes states with different partial waves, such that one should replace $L$ = 1 with the lowest centrifugal barrier associated with the odd-$L$ adiabatic channel. Similarly, $L$ = 0 becomes the lowest-energy adiabatic channel with even $L$, which does not have a centrifugal barrier.

\begin{figure}
\begin{center}
\includegraphics[width=9cm]{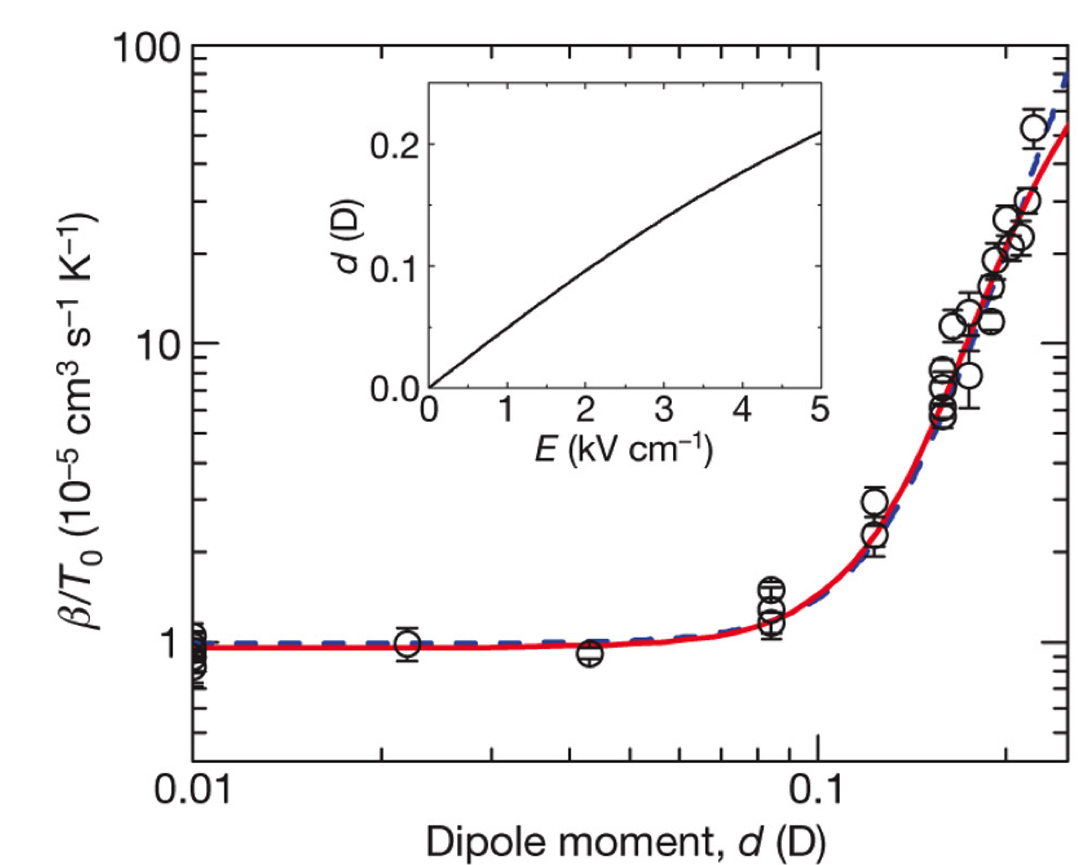}
\caption{Decay coefficient as a function of induced dipole moment in a bulk 3D gas. A clear onset of increased loss at ~0.1 D is apparent. The inset shows the induced dipole moment as a function of electric field. Reproduced with permission from Reference~\cite{Ni2010}.}
\label{fig:dipdep}
\end{center}
\end{figure}

In a regular far-off-resonance optical dipole trap, we do not have control over the projection of the relative angular momentum of the colliding particles on the quantization axis. Therefore, while for $m_L= \pm 1$ the centrifugal barrier is significantly increased with a reasonable electric field, the observed loss rate will be dominated by the $m_L=0$ collision channel, which has a much lowered barrier height for reactions (Fig.~\ref{fig:pwave}).  Consequently, the measured loss rate increases sharply as a function of the electric field, as shown in Fig.~\ref{fig:dipdep}.  In fact, $\beta \propto d^6$ ($d$ is the induced dipole moment in the laboratory frame), which agrees well with a quantum threshold model that takes into account the long-range dipolar interaction~\cite{Ni2010}. The dependence of the rate constant on electric field shows a smooth behavior, further indicating a near-unity reaction probability in the short range.

Chemical reactions for ultralow temperature polar molecules confined in a three-dimensional harmonic trap thus depend strongly on the quantum statistics of the molecular gas, individual collision partial waves and temperature, and the induced dipole moment in the lab frame. In the remainder of this Chapter we will present results on the control of the reaction rate when we modify the dimensionality of the optical traps in which the molecules are confined.

\subsection{Inducing the dipole moment with DC and AC electric fields}

\begin{figure}
\begin{center}
\includegraphics[width=10cm]{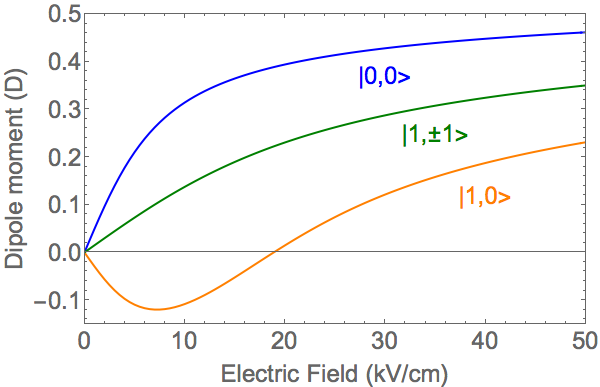}
\caption{Induced dipole moments of KRb for the $\ket{\tilde{0},0}$ (blue), $\ket{\tilde{1},0}$ (orange), and $\ket{\tilde{1},\pm 1}$ states (green), calculated using the lowest 20 rotational levels in Eq.~\ref{eqn:rrham}.}
\label{fig:dipolemoments}
\end{center}
\end{figure}

Since an electric dipole moment exists only between quantum states that have opposite parity, individual rotational states $\ket{N,m_N}$ have no dipole moment in the lab frame at zero electric field.  When a DC electric field is applied, opposite-parity rotational states become mixed, giving rise to a nonzero dipole moment in the lab frame. In other words, molecules become aligned with the applied field.  The Hamiltonian for the rotational states $\ket{N,m_N}$ under an electric field ($\boldsymbol{\epsilon}$) has two contributions arising from the diagonal rotational energy and the state-mixing (off-diagonal) Stark effect:
\begin{equation}\label{eqn:rotcoup}
\bra{N,m_N} \hat{H} \ket{N',m_N'} = B N (N+1) \delta_{N,N'} \delta_{m_N, m_{N'}} -\bra{N,m_N} \hat{\mathbf{d}} \cdot \boldsymbol{\epsilon} \ket{N',m_N'}.
\end{equation}
Here $B$ is the rotational constant for the vibrational ground state, and the term $-\hat{\mathbf{d}} \cdot \boldsymbol{\epsilon}$ describes the dipole-field coupling operator.  If the electric field, of magnitude $\epsilon$, is along the $z$ axis, the matrix elements of Eq.~\ref{eqn:rotcoup} simplify to
\begin{multline}\label{eqn:rrham}
\bra{N,m_N} \hat{H} \ket{N',m_N'} = B N (N+1) \delta_{N,N'} \delta_{m_N, m_{N'}}  - \\
 \mathcal{D} \epsilon  \sqrt{(2N+1)(2N'+1)}(-1)^{m_N}
\left( \begin{array}{ccc}
N & 1 & N'  \\
-m_N & 0 & m_N'
 \end{array} \right)
 \left( \begin{array}{ccc}
N & 1 & N'  \\
0 & 0 & 0
 \end{array} \right).
\end{multline}
Here $\mathcal{D}$ is the permanent dipole matrix element (0.574 Debye for KRb) and the terms in parentheses are 3-$j$ symbols.  The first 3-$j$ symbol is zero unless $m_N = m_{N'}$. So the eigenstates in an electric field are
\begin{equation}\label{eqn:dressedsum}
\ket{\tilde{N},m_N}= \sum_{N'} c_{N'} \ket{N',m_N},
\end{equation}
where $\ket{\tilde{N},m_N}$ is the state that adiabatically connects to $\ket{N,m_N}$ when the field is turned down to zero. The dressed state energies $E_{\tilde{N},m_N}$ can be determined by diagonalizing Eq.~\ref{eqn:rrham}, and then the induced dipole moment of state $\ket{\tilde{N},m_N}$ is known through the quantity, $- \frac{\partial E_{\tilde{N},m_N}}{\partial \epsilon}$.  The induced dipole moments for the $N=0$ and $N=1$ states of KRb are displayed in Fig.~\ref{fig:dipolemoments}.  A field strength of more than 40 kV/cm in the lab is required to reach 80\% of the full dipole moment. In general, the value of this critical field scales as $B/\mathcal{D}$, and it can be much smaller for some bialkali molecular species.

Since successive rotational states have opposite parity, a more efficient option for realizing a lab-frame dipole moment is to use resonant RF or microwave field to directly couple two neighboring rotational states. When the $N=0$ and $N=1$ states are coherently mixed, we realize a dipole moment of the magnitude of $\mathcal{D}/\sqrt{3}$ even after the applied field has been removed. This is the technique used in References~\cite{Yan2013, Hazzard2014} to realize dipolar-interaction-mediated spin exchanges in a three-dimensional optical lattice.

\subsection{Reduced dimensions: quantum stereodynamics of chemical reactions in a 2D gas}

To suppress the large reaction rate of KRb molecules in an electric field, we must remove the lossy head-to-tail collisions. Hence, we need to gain control over the projection of the relative angular momentum of the colliding molecules on the quantization $z$ axis, $m_L$. This can be achieved by confining molecules in a two-dimensional optical trap, which can be realized by interfering two counterpropagating optical beams to create a 1D lattice that consists of a stack of 2D lattice sites. The DC electric field is applied perpendicular to the 2D traps. In such a configuration, the lossy ``head-to-tail" collisions with $m_L = 0$ are greatly suppressed, and the repulsive ``side-by-side" collisions with $m_L = \pm 1$ stabilizes the molecular gas~\cite{Miranda2011, micheli2007, buchler2007}.

If a strong optical confinement along the electric field direction is achieved, the motion along this axis will be fully quantized, labelled with a quantum number $\nu_z$. At ultralow temperatures, the majority of molecules will occupy the motional ground state of $\nu_z = 0$, with only a small fraction of them in $\nu_z = 1$.  The angular momentum $L$ in the 3D case is no longer a good quantum number in 2D; instead $\nu_z$ and the angular momentum projection $M$ along $z$ become relevant in the antisymmetrization of the two-molecule wavefunction for fermionic KRb. For example, if two KRb molecules are prepared in the same internal state (spin-polarized) and have the same $\nu_z$, then they must collide with an odd $M$, which will give rise to a repulsive dipole-dipole interaction that suppresses the chemical reaction. If, on the other hand, the two molecules are prepared in different states of $\nu_z$, then the even $M$ channel will be left open, and the reaction will thus proceed strongly. In the limit of a very large applied electric field and a very strong optical confinement along $z$, we will approach the classical limit where repulsive dipole-dipole interactions dominate for both Bose and Fermi quantum statistics~\cite{Miranda2011, micheli2007, buchler2007}. This is the regime where the characteristic length scale associated with the dipole-dipole interaction exceeds the harmonic oscillator length in $z$.

\begin{figure}
\begin{center}
\includegraphics[width=14cm]{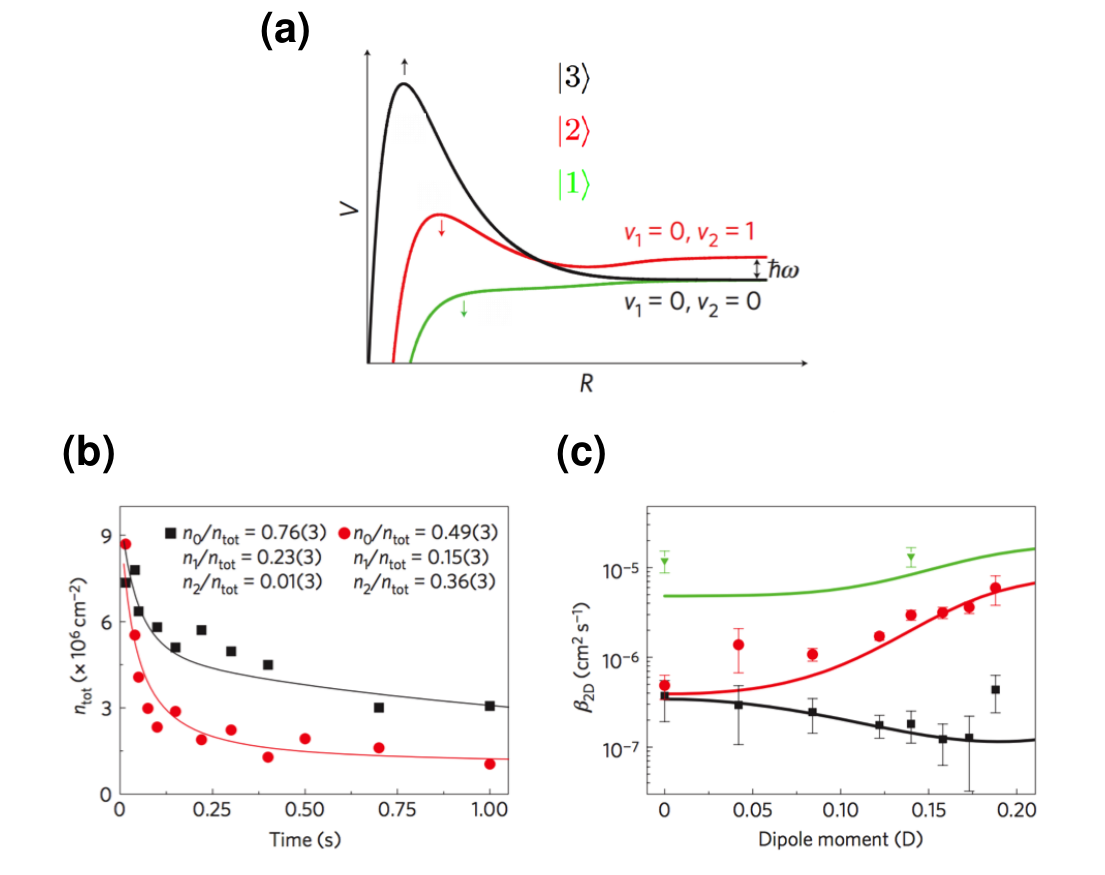}
\caption[Chemical reactions in the 1D lattice]{Chemical reactions in the 1D lattice.  (a) The barrier for collision channel 1 (green, distinguishable molecules), 2 (red, indistinguishable molecules in different bands), and 3 (black, indistinguishable molecules in the same band).  Channel 3 is the most desirable to stabilize the molecular gas in a 2D geometry since it corresponds to repulsive side-by-side collisions.  The arrows next to the curves indicate whether the barrier increases or decreases when the electric field increases.  (b) Loss curves for two different initial band populations.  The initial loss rate is faster when a larger fraction of molecules occupy higher bands initially.  (c) The measured 2D loss coefficients $\beta_\text{2D}$ for the three cases depicted in (a), with the same color coding.  Reproduced with permission from Reference~\cite{Miranda2011}.}
\label{fig:cr2d}
\end{center}
\end{figure}

This interesting case of stereo-chemical reactions in the quantum regime was experimentally studied in Ref.~\cite{Miranda2011}, where the reaction rate as a function of the molecular temperature, the harmonic confinement along $z$, and the induced dipole moment was systematically measured. Specifically, Fig.~\ref{fig:cr2d}a documents the three different configurations that were studied in the experiment~\cite{Miranda2011}: (1) molecules occupy different internal states, (2) spin-polarized molecules with different $\nu_z$, and (3) spin-polarized molecules in the same state of $\nu_z$. Collisions via Channel 1 are $s$-wave and result in very fast loss.  Channel 2 is also undesirable, since it corresponds to unsuppressed ``head-to-tail" collisions with even $M$.  Channel 3 is the most ideal case, as it corresponds to ``side-by-side" collisions with odd $M$.  Parametric heating was used to promote molecules to higher bands, which sets the relative importance of channels 2 and 3.  Figure~\ref{fig:cr2d}b displays measured loss curves for two different initial conditions of molecular temperature. The black curve shows the loss for a low temperature distribution, where about 75\% of the molecules were initially in the lowest band. The red curve shows the loss for a distribution where parametric heating was applied to the gas and a larger fraction of molecules was promoted to the second excited band. The decay curves are used to extract the loss rate coefficient $\beta_\text{2D}$ for the 3 channels (Fig.~\ref{fig:cr2d}c).  As expected, $\beta_\text{2D}$ is the smallest for channel 3.  The suppression of the reaction rate via odd $M$, in comparison to the head-to-tail collisions, is about a factor of 60 at an induced dipole moment of 0.174 D (Fig.~\ref{fig:cr2d}c).

This two-dimensional trap configuration can be ubiquitously applied to dipolar systems to control collision processes by fine tuning the collision angle relative to the dipole angle. Similar experiments have recently been performed with Feshbach molecules of highly magnetic Erbium atoms confined in a 1D lattice with similar goals of removing the attractive part of the dipole-dipole interaction~\cite{Frisch2015}. New experiments at MIT with NaK molecules are also moving in the direction of a 1D lattice to use strong repulsion to stabilize the system against inelastic collisions~\cite{ZwierleinPC}.

\subsection{Sticky collisions: 3-molecule collision channels}
The previous discussion of molecular collisions and reactions focused on two-body inelastic collisions. This is strongly supported by the observation that inelastic loss rates are $\sim$100 times larger with non-identical fermions than with spin-polarized fermions. To reiterate, in the spin polarized case the molecules have to collide in the $p$-wave channel, whereas with a spin mixture the $s$-wave channel becomes available and so the molecules do not have to tunnel through the energy barrier of the $p$-wave channel. Given that the KRb + KRb reaction is exothermic, there is a strong interest in the scientific community in working with bialkali species that are predicted to have endothermic reactions with suppressed loss rates~\cite{Zuchowski2010,Quemener2011,Byrd2012}.  However, molecular scattering experts such as John Bohn at JILA have also predicted substantial loss in three-body collision channels~\cite{Croft2014, Mayle2012, Mayle2013, Frisch2014}. The simple intuitive picture they invoke for these processes is the following: two molecules can collide and temporarily form a reaction complex prior to separating. During the period in which they are very near each other, a third molecule can come in and force the complex to a deeper bound molecular state that results in inelastic loss of all three molecules. These processes could be responsible for observed losses that are difficult to attribute to two-body loss mechanisms, but rigorous experimental confirmation of these lossy three-body processes is challenging.

Let us consider the case of three-body collisions of fermions. If the fermions are spin polarized, three-body collisions will require the $p$-wave channel in the same way as two-body collisions. However, if the molecules are not spin polarized, the difference between two-body and three-body could become experimentally distinguishable. For example, suppose we restrict ourselves to only two unique internal states for the fermions. Then, collisions between two non-identical fermions can take place in the $s$-wave channel, and would not require tunneling through a barrier. However, three-body collisions will necessarily still require a $p$-wave channel since two of the three fermions must be identical. In this situation, therefore, we would not expect the three-body loss rate to be very different between identical and non-identical fermions. This is in stark contrast to the scenario we have presented on the KRb loss, which again points to two-body loss as the dominant loss process. The work at MIT with NaK molecules~\cite{ZwierleinPC, Will2016}, which are ostensibly not reactive in the two-body channel, could shed light on three-body loss.

With bosonic molecules the situation becomes significantly more complex. Since bosons can collide in the $s$-wave channel, enormously large loss can happen for two-body collisions and three-body collisions. Even bosonic molecules that are ostensibly chemically stable have been observed to decay very quickly~\cite{Takekoshi2014, Molony2014}, and experimental signal-to-noise ratios make it very difficult to differentiate between two-body and three-body loss mechanisms. Just as for our reactive KRb molecules, these experiments can benefit from confining the molecules in zero spatial dimension, achieved with a three-dimensional optical lattice, to completely remove collisional losses. We will thus focus on the production, stabilization, and applications of ultracold polar molecules in 3D optical lattices in the next Section.

\section{Suppression of chemical reactions in a 3D lattice}
For many experiments based on the long-range interactions of polar molecules, contact between the molecules in unnecessary or even unwanted. For such experiments, a full 3D optical lattice can be used to pin molecules to individual sites where they will remain without colliding for the duration of the experiment. While research on ultracold polar molecules in optical lattices had started only recently~\cite{Danzl2010, Chotia2012}, promising initial experimental results have already been achieved, as will be summarized in this Section. We begin by discussing the case of a weak optical lattice, where tunneling between sites and the resulting collisions between molecules can be investigated. We will consider the case of a spin-mixture that is allowed to tunnel in the lattice, which gives rise to a manifestation of the so-called continuous quantum Zeno mechanism. With an understanding of these effects as well as the role of quantum statistics on stabilizing molecules in an optical lattice, we can then investigate the case of a spin-polarized gas of molecules in a three-dimensional lattice where very long lifetimes are observed. We then proceed to describe several very interesting experiments that were performed in such a configuration.

\subsection{The continuous quantum Zeno mechanism: stability from strong dissipation}

The largest measured two-body loss rate for KRb at zero electric field was for the rotational mixture.  In this case, $s$-wave collisions are allowed and the on-site loss rate in a 3D lattice is very large and can be larger than both the tunneling rate and the energy gap between the two lowest lattice bands.  In this regime, the process for a molecule to tunnel onto a lattice site that is already occupied by another molecule is greatly suppressed by the continuous quantum Zeno effect~\cite{Syassen2008}, and, somewhat counterintuitively, the effective loss rate of the whole system decreases as the on-site loss rate increases.

\begin{figure}
\begin{center}
\includegraphics[width=14cm]{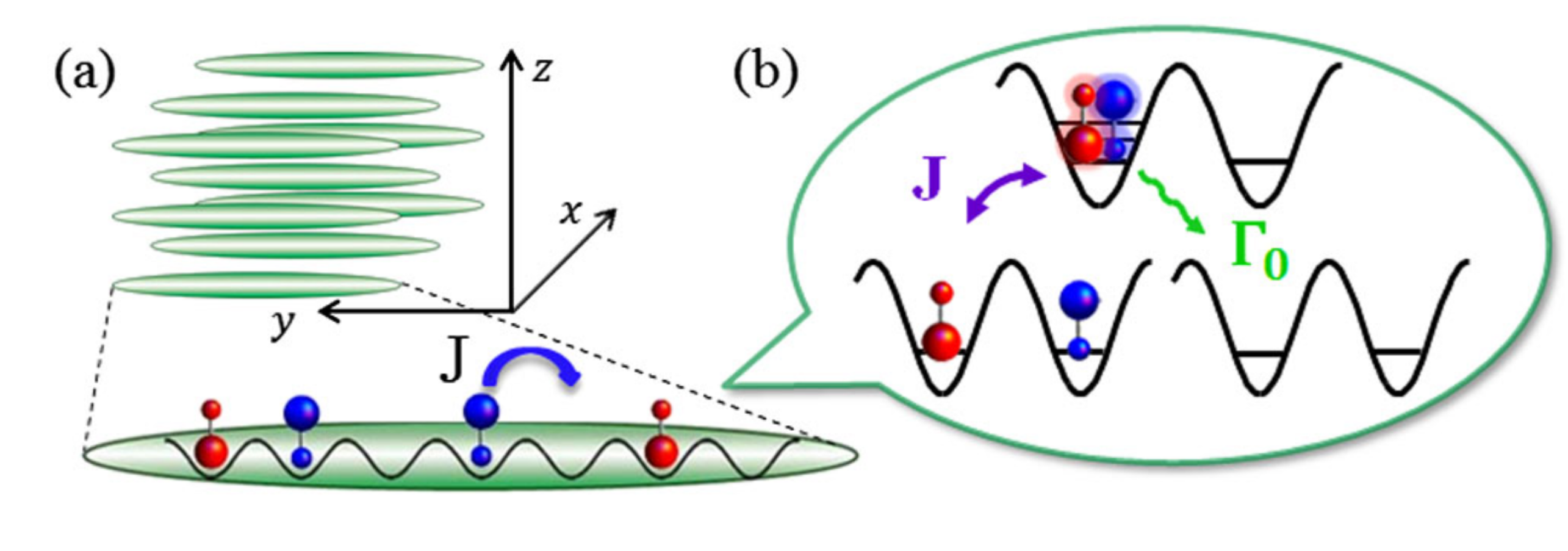}
\caption[Experimental setup for studying the continuous quantum Zeno effect]{(a) Experimental setup for studying the continuous quantum Zeno effect.  The molecules are held in a 3D lattice, with strong lattices along $x$ and $z$ and a weak lattice along $y$.  This realizes a system of decoupled 1D tubes with a weak corrugation along the tubes.  (b) The tunneling energy along the tube direction is $J$ and the onsite loss rate is $\Gamma_0$.  Reproduced with permission from Reference~\cite{Zhu2014}.}
\label{fig:zenosetup}
\end{center}
\end{figure}

The experimental setup for the work reported in Refs.~\cite{Zhu2014, Yan2013} is shown schematically in Fig.~\ref{fig:zenosetup}. About $10^4$ ground-state molecules were produced in a deep 3D lattice (along all three directions).  An incoherent mixture of $\ket{\da} = \ket{0,0,-4,1/2}$ and $\ket{\ua} =\ket{1,-1,-4,1/2}$ was created by first applying a $\pi/2$ pulse between these two rotational states and then waiting 50 ms for decoherence to set in. The two rotational states constitute a pseudo-spin-1/2 system.  Then, the lattice depth along $y$ was reduced within 1 ms to allow tunneling at a rate $J_y/h$.  This creates a system of decoupled 1D tubes with a weak lattice along the tube axis ($y$).  After holding the molecules in the lattice for a variable amount of time, the number of molecules in the $N=0$ state ($\ket{\da}$) was measured to determine the loss of this mixed-spin system.

There is no Pauli suppression preventing a $\ket{\da}$ and a $\ket{\ua}$ molecule from occupying the same lattice site, and their on-site loss rate $\Gamma_0$ is very large.  In fact, $\Gamma_0 \gg \frac{J_y}{\hbar}$. In this limit, tunneling-induced on-site loss becomes a second-order perturbative process and the effective loss rate for the overall system becomes
\begin{equation}\label{eqn:effloss}
\Gamma_{\text{eff}}=\frac{2 J_y^2}{\hbar^2 \Gamma_0}.
\end{equation}
Since particles are lost via a two-body process,
\begin{equation}\label{eqn:zenoloss}
\dot{N_i}(t)= -\frac{\kappa N_i(t)^2}{N_i(0)},
\end{equation}
where $i = \ket{\ua}$ or $\ket{\da}$, $\kappa = 4 q \Gamma_\text{eff} n_{\da}(0)$, and $N_i(0)$ is the initial number of molecules in state $i$.  $n_{\da}(0)$ is the initial filling fraction of $\ket{\da}$ molecules, so $2 n_{\da}(0)$ is the total filling fraction, and $q$ is the number of nearest neighbors.  Measuring the dependence of $\kappa$ on $J_y$ and $\Gamma_0$ is thus directly achievable in the experiment. An extra benefit is the determination of the molecule filling fraction in the 3D lattice.

\begin{figure}
\begin{center}
\includegraphics[width=17cm]{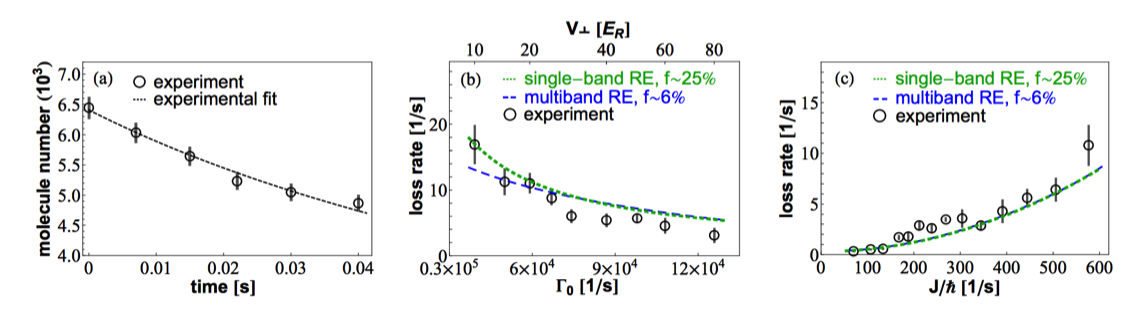}
\caption[Extracting the filling fraction from the loss measurements]{(a) A typical loss curve, where the data is fit to the solution of the rate equation (Eq.~\ref{eqn:zenoloss}).  (b) and (c) The dependence of $\kappa$ on $\Gamma_0$ (b) and $J_y\hbar$ (c).  See the text for more details.  Reproduced with permission from Reference~\cite{Zhu2014}.}
\label{fig:zenodep}
\end{center}
\end{figure}

The transverse lattice depth $V_\perp$ was typically $40 \, E_R$, while the lattice along $y$ direction was varied between 5 and 16 $E_R$. Here, $E_R$ is the lattice photon recoil energy for KRb. To vary $\Gamma_0$ only, the transverse lattice depths were changed. To vary $J_y$, the lattice depth along the $y$ direction was changed; however, to keep $\Gamma_0$ fixed, the transverse lattice depths were also changed accordingly. Figure~\ref{fig:zenodep}a shows a typical loss curve for $V_y = 5 E_R$ and $V_\perp = 25 E_R$.  The data is fit to the solution of Eq.~\ref{eqn:zenoloss}.  Na\"{i}vely calculating the filling using Eq.~\ref{eqn:zenoloss} provides an overestimate of the filling fraction due to the fact that the molecular population in higher lattice bands is not taken into account in this equation. Since $\Gamma_0$ is on the same order of magnitude as the band gap, higher band populations must be included to reach the correct value of $\Gamma_0$.  For the experimental parameters, properly accounting for higher bands can decrease $\Gamma_0$ by about a factor of 5, which then increases $\Gamma_\text{eff}$ by the same factor.  The curve in Fig.~\ref{fig:zenodep}b clearly demonstrates that $\kappa$ decreases as $\Gamma_0$ increases, while the data in Figure~\ref{fig:zenodep}c shows that $\kappa$ has roughly a quadratic dependence on $J_y/\hbar$. These results constitute the strong evidences for the continuous quantum Zeno effect, and at the same time we determine the lattice filling fraction to be around 6\%.

\subsection{Long-lived molecules in a 3D optical lattice}
Two years before the production of ground-state polar molecules, researchers had already prepared Feshbach molecules in an optical lattice by magneto-association of pairs of atoms contained in single sites of a bosonic Mott insulator of Rb~\cite{Thalhammer2006}. They were then able to remove unpaired atoms, leaving only sites that were either empty or contained a Feshbach molecule. In such an experiment, no collisions between molecules are possible and therefore molecules cannot undergo inelastic scattering via vibrational quenching from the Feshbach state. The lifetime was measured to be 700 ms, which is significantly longer than that in a bulk gas where molecules can collide with atoms and with each other~\cite{Thalhammer2006}. However, there is no fundamental limit to this lifetime, and it is ultimately limited by inelastic light scattering from the trapping light.

More recently, we have demonstrated the creation of KRb Feshbach and ground-state molecules in 3D optical lattices, and have shown that their lifetimes can exceed 20 seconds~\cite{Chotia2012}. This lifetime is also limited by off-resonant light scattering from the optical trap, which is fairly far detuned from molecular resonances. The lifetime of these molecules in the optical lattice is significantly longer than the timescales that are relevant for the dominant interaction energies between the molecules~\cite{Lemeshko2012, Hazzard2013}, and so this offers an excellent starting point for the investigation of dipole-dipole interactions in an optical lattice and for the establishment of a lattice model with a spin-1/2 Hamiltonian.

As discussed above, applying an electric field in the laboratory frame can potentially reduce the lifetime of the molecular gas by allowing the attractive dipolar interaction to reduce the $p$-wave barrior. In the case of a deep 3D lattice and therefore the absence of tunneling, the short-range collisional physics is fully suppressed, and hence the mechanism for reducing the lifetime by an electric field is no longer valid. Indeed, we observe experimentally that the lifetime of the molecules in a deep 3D lattice remains the same with an electric field of $\sim 4$ kV/cm as with no field. This is very encouraging: we can now turn on the full-strength dipole-dipole interactions with electric fields without having to worry about any extra loss our molecular sample may experience.

\begin{figure}
\begin{center}
\includegraphics[width=8cm]{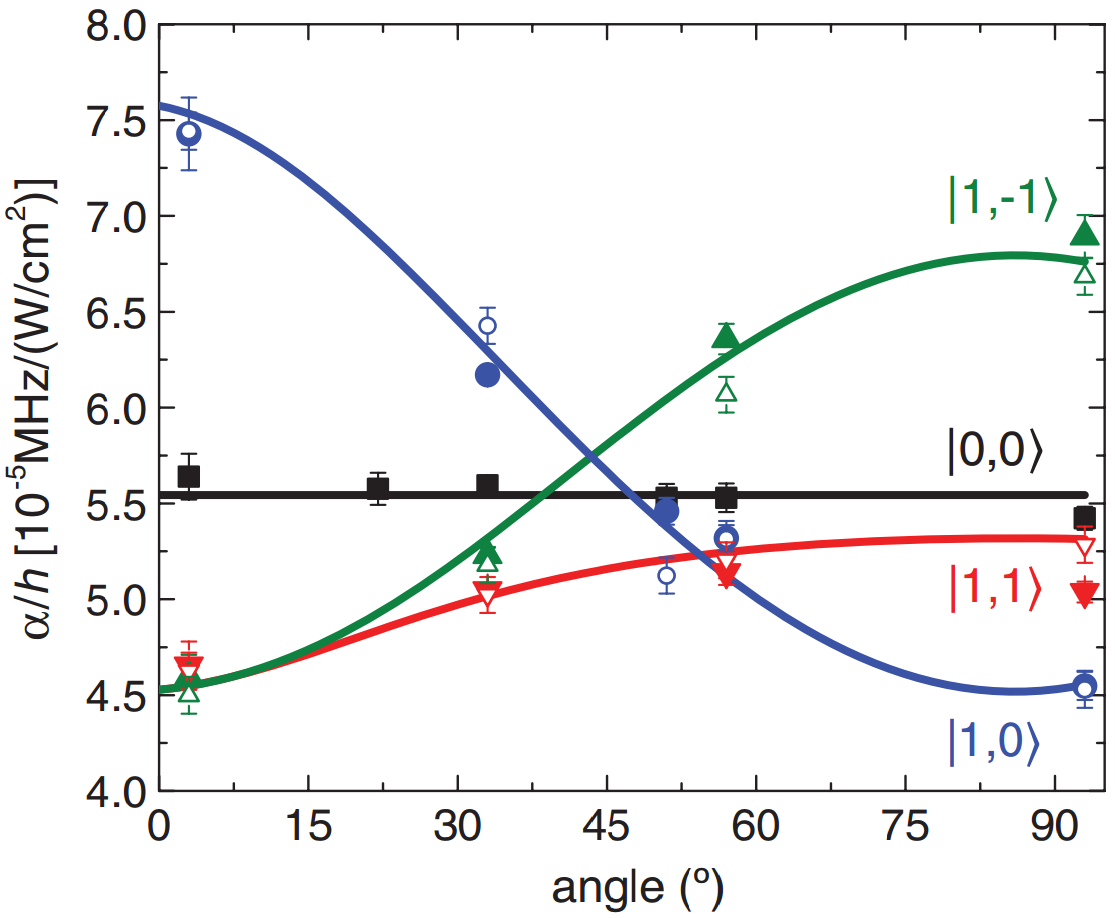}
\caption[The magic angle]{The ac polarizability of KRb at 1064 nm for the $\ket{0,0}$ (black squares), $\ket{1,0}$ (blue circles), $\ket{1,1}$ (red inverted triangles), and $\ket{1,-1}$ (green triangles) states as a function of the angle between the quantization field (B-field) and the polarization of the light field in a 1D optical lattice. Reproduced with permission from Reference~\cite{Neyenhuis2012}.}
\label{fig:AnPol}
\end{center}
\end{figure}

Moreover, a long lifetime in the lattice allows for investigations of the real part of the molecular ac polarizability in addition to its lossy imaginary component. This is important for us to understand the molecular energy level shifts inside the lattice. The wavefunctions for many molecular states are spatially anisotropic, and hence the corresponding polarizability depends on the angle between the polarization of the lattice light and the quantization field defined by an applied electric or magnetic field. By tuning this angle carefully, KRb molecules are found to exhibit a `magic' trapping condition in which the polarizability of two rotational states (as used to define a spin-1/2 system) can be adjusted to match with each other~\cite{Neyenhuis2012} (see figure~\ref{fig:AnPol}). Such `magic' conditions are reached by adjusting the lattice light wavelength in optical lattice clocks and are routinely used to minimize the effects of the trapping light on the clock transition~\cite{Ludlow2015}. For molecules in this `magic' angle lattice, the coherence time for a coherent superposition of the two rotational states is found to be orders of magnitude longer than otherwise, which is an important prerequisite for studying spin physics with polar molecules in an optical lattice. We note that a coherence time of roughly one second has been observed recently between two hyperfine states of NaK in a single rotational state~\cite{Park2016}.

\subsection{Quantum magnetism with polar molecules in a 3D optical lattice}

\begin{figure}
\begin{center}
\includegraphics[width=6cm]{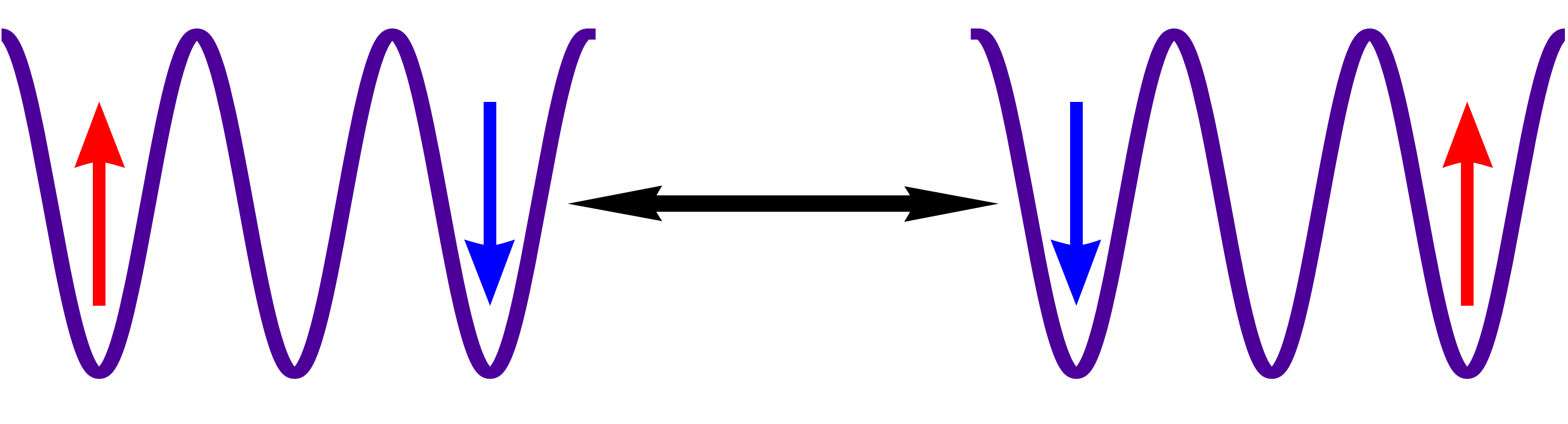}
\caption[Schematic of spin exchange]{Spin exchange between molecules in $N=0$ ($\da$) and $N=1$ ($\ua$).  Because of the long-range interactions, the exchange can occur in principle over arbitrary distances in the lattice.}
\label{fig:exchange}
\end{center}
\end{figure}

With chemical reactions under control and the molecular gas stabilized in the 3D optical lattice, we have recently begun to study a conservative manifestation of the dipole-dipole interaction, where the molecules never come in contact with each other. This is a new experimental regime where the internal degrees of freedom (spins) of molecules interact strongly, but the motional degrees of freedom are largely decoupled from the system dynamics. Certainly this is not the case for cold atoms in a lattice that rely on short-range interactions.  References~\cite{Barnett2006, gorshkov2011} proposed that polar molecules can be used to study quantum magnetism, and dipolar interactions should be observable even at low lattice fillings and high entropies~\cite{Hazzard2013}.  Specifically, by encoding a spin-$1/2$ degree of freedom in the lowest two rotational states ($\ket{\da} \equiv \ket{0,0}$ and $\ket{\ua} \equiv \ket{1,-1}$ or $\ket{\ua} \equiv \ket{1,0}$), the molecules can undergo energy-conserving spin exchanges mediated by dipolar interactions between nearby lattice sites, as depicted in Fig.~\ref{fig:exchange}.
The Hamiltonian that describes this exchange process is~\cite{gorshkov2011,Yan2013,Wall2015}:
\begin{equation}\label{eqn:goal}
\hat{H} = \frac{1}{2} \sum_{i \neq j} V_{dd}(\mathbf{r_i}-\mathbf{r_j}) \left( \frac{J_\perp}{2} \left(\hat{S}_i^{+} \hat{S}_j^{-} + \hat{S}_i^{-} \hat{S}_j^{+}  \right) \right),
\end{equation}
where $V_{dd}(\mathbf{r_i}-\mathbf{r_j})= \frac{1-3 \cos^2 \theta_{ij}}{r_{ij}^3}$, which is a geometrical factor pertaining to dipole-dipole interactions, and $\hat{S^+}$ and $\hat{S^-}$ are spin-$1/2$ raising and lowering operators.  The coupling constant $J_\perp = \frac{k d_{\da \ua}^2}{4 \pi \epsilon_0 a_\text{lat}^3}$, where $d_{\da \ua}$ is the transition dipole moment between the two spin states $\ket{\da}$ and $\ket{\ua}$ (at zero DC field, $d_{\da \ua} = \mathcal{D} / \sqrt{3}$), $\epsilon_0$ is the permittivity of free space, $a_\text{lat}$ is the lattice spacing, and subscripts $i$ and $j$ index lattice sites. We also note that $k=2$ for $\ket{\ua} = \ket{1,0}$ and $k=-1$ for $\ket{\ua} = \ket{1,-1}$, which are set by the matrix elements coupling to $\ket{\da} \equiv \ket{0,0}$.  With molecules distributed in a 3D lattice, the dominant frequency (energy) component present in the Hamiltonian is $|J_\perp/(2 h)|$, where $h$ is Planck's constant.  The experiments of References~\cite{Yan2013, Hazzard2014} were performed at zero DC electric field; at nonzero field, Equation~\ref{eqn:goal} has additional terms including an Ising interaction $\sim \sum_{i,j} J_z \hat{S_i^z} \hat{S_j^z}$.

\begin{figure}
\begin{center}
\includegraphics[width=16cm]{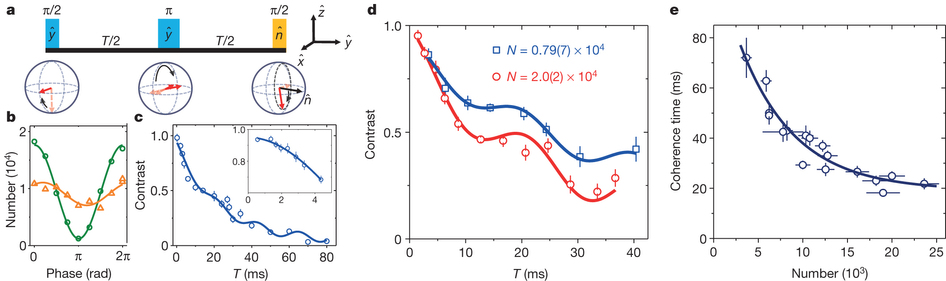}
\caption{(a) Spin echo setup and timing diagram. (b) Typical Ramsey fringes.  (c) Contrast vs.~time showing the decay and oscillations.  The inset shows that the contrast decay for short times is concave down.  (d) Contrast decay for two different particle densities.  The oscillation frequency is basically the same for the two datasets but the coherence time is shorter for higher density.  (e) The data is fit to $C(T) = A e^{-T/\tau}+B \cos^2 (\pi f T)$.  Compiling all of the measured coherence times shows $\tau \propto 1/N$.  Reproduced with permission from Reference~\cite{Yan2013}}
\label{fig:bofig2}
\end{center}
\end{figure}

The dynamics are probed with a spin-echo Ramsey spectroscopy sequence, which is schematically depicted in Fig.~\ref{fig:bofig2}a. One additional complication in the experiment is that there is a site-to-site differential energy shift due to a residual inhomogeneous light shift arising from the molecules' anisotropic polarizability. The use of a 'magic' angle for the lattice polarization with respect to the quantization axis reduces the inhomogeneity by more than a factor of 10, but a residual effect remains~\cite{Neyenhuis2012}.  A spin echo technique is an effective approach to mitigate the effects of single-particle dephasing from this energy spread across the molecular cloud. Initially, all of the molecules are $\ket{\da}$.  The first $\pi/2$ pulse excites every molecule to a coherent superposition of $(\ket{\ua}+\ket{\da})/\sqrt{2}$.  After a free evolution time $T/2$, an echo pulse is applied to reverse the single-particle dephasing from the inhomogeneous light shift.  After another free evolution time of $T/2$, a final $\pi/2$ pulse is applied with a phase $\phi$ relative to the first pulse.  This rotates the Bloch vector about the axis $\hat{\mathbf{n}} = \cos \phi \, \hat{\mathbf{y}}+\sin \phi \, \hat{\mathbf{x}}$. We measure the number of molecules in $\ket{\da}$ by using STIRAP to transfer the ground-state molecules back to weakly bound Feshbach molecules, followed by the normal atomic absorption imaging protocol. By varying the angle $\phi$ we can obtain a Ramsey fringe contrast for a given free evolution time $T$, as shown in Fig.~\ref{fig:bofig2}b, which we fit to $\frac{N_\text{tot}}{2} \big( 1+C \cos(\phi+\phi_0) \big)$. Here, $N_\text{tot}$ is the total molecule number and $C$ is the fringe contrast ($ 0 \le C \le 1$).  The contrast determines the amount of spin coherence left in the system after a time $T$, and it usually decays as a function of $T$ due to residual single-particle dephasing and many-molecule interaction effects. Some typical contrast decay curves are shown in Figure~\ref{fig:bofig2}c and d.  The most striking feature of these curves is the oscillation superimposed on an overall decay.  We attribute both the oscillations and the overall decay to dipolar interactions.  Dilute lattice fillings and long-range interactions lead to a spread of interaction energies, which cause dephasing and loss of contrast, and the interaction energy spectrum has the strongest contribution from the nearest-neighbor interaction (of frequency $J_\perp/(2 h)$).

Furthermore, the curves in Fig.~\ref{fig:bofig2}d display a clear density dependence, which is a signature of an interaction effect. The density was varied by loading the same initial distribution of molecules and then holding them in the lattice for a variable amount of time to allow for single-particle loss from light scattering. Thus, the density is roughly proportional to the number of molecules we detect~\cite{Chotia2012}.  We expect $\tau \propto 1/N$, since
\begin{equation}
\tau \propto \frac{1}{\langle E_\text{int} \rangle} \propto \frac{\bar{R}^3}{J_\perp} \propto \frac{1}{J_\perp \rho},
\end{equation}
where $\bar{R}$ is the average interparticle spacing and the density $\rho = \bar{R}^{-3}$.  For our loading scheme, $\rho \propto N$.  Figure~\ref{fig:bofig2}e shows the coherence time clearly scales as $1/N$.

\begin{figure}
\begin{center}
\includegraphics[width=14cm]{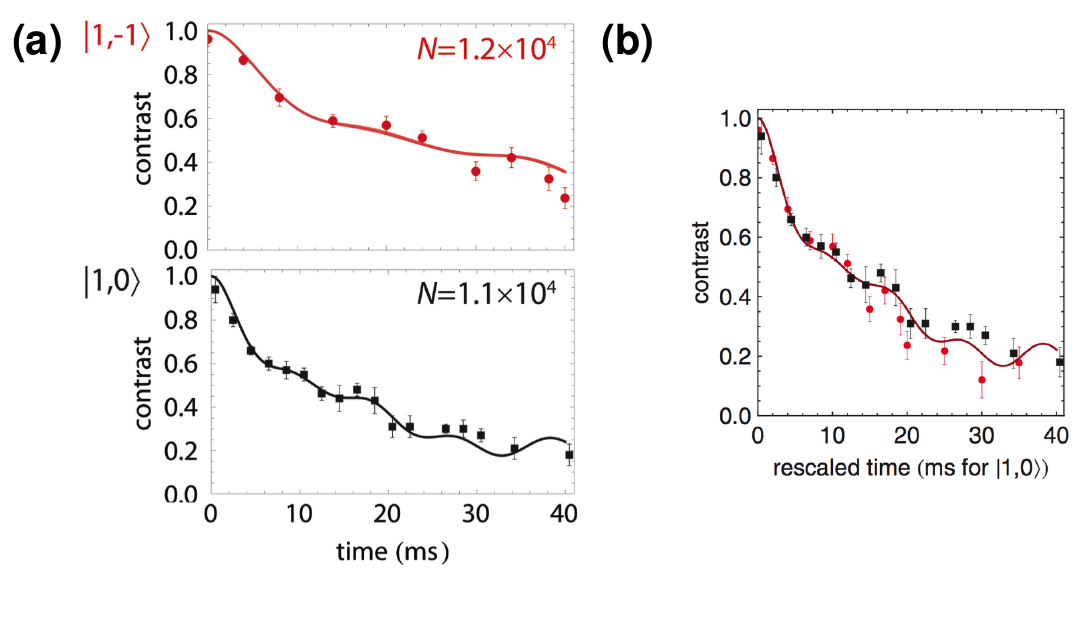}
\caption[Comparing contrast decay for $\ket{\ua} = \ket{1,-1}$ and $\ket{\ua}=\ket{1,0}$]{(a) Typical contrast decay curves for $\ket{\ua}= \ket{1,-1}$ (top) and $\ket{\ua}=\ket{1,0}$ (bottom) for roughly the same density ($\sim 1.2 \times 10^4$ molecules).  The coherence time is clearly shorter for the $\ket{1,0}$ data, which is expected due to the stronger interactions.  (b) Taking the two datasets from (a) and scaling the time axis for the $\ket{1,-1}$ data by a factor of two, we see that the curves collapse onto each other reasonably well, which highlights that dipolar interactions are responsible for the observed dynamics. The lines are theory~\cite{Hazzard2014}. Reproduced with permission from Reference~\cite{Hazzard2014}.}
\label{fig:101m1comp}
\end{center}
\end{figure}

Another way to change the interaction strength is to couple $\ket{0,0}$ to a different state in the $N=1$ manifold, as $J_\perp$ is twice as large for the $\{ \ket{0,0}, \ket{1,0} \}$ transition.   To test this, the same experiment described above was repeated with $\ket{\ua} = \ket{1,0}$~\cite{Hazzard2014}.  Figure~\ref{fig:101m1comp}a shows typical contrast decay curves for $\sim10^4$ molecules for both choices of $\ket{\ua}$, and Figure~\ref{fig:101m1comp}b compares the two curves when time is rescaled by a factor of 1/2 for the $\ket{1,-1}$ data.  The contrast clearly decays faster for $\ket{\ua} = \ket{1,0}$ for which the interactions are stronger, and the fact that the two curves nearly collapse onto each other when time is rescaled by exactly the same ratio of the interaction energy highlights that dipole-dipole interactions give rise to the observed dynamics.  The solid curves are based on theoretical simulations obtained from a cluster expansion~\cite{Hazzard2014}.  A new ``moving-average" cluster expansion was developed in order to explain our data, which exemplifies the need of new development of theory tools to help understand some of the complex many-body quantum systems that are now running in laboratories.  This theory comparison also informed that the molecular filling fraction in the lattice was low, at about 5\% for the data in Fig.~\ref{fig:101m1comp}, which is consistent with the earlier estimation based on the quantum Zeno effect. This has motivated work to increase the filling fraction in the lattice~\cite{Moses2015}.

\subsection{A low entropy quantum gas of polar molecules in a 3D optical lattice}

The first successful creation of KRb ground state molecules in 2008 reached close to the quantum degeneracy, with a temperature of 1.3$T_F$, where $T_F$ is the Fermi temperature. However, despite an intense effort to further cool the molecular gas, the lowest temperature for the KRb gas reached in a harmonic trap is $T/T_F =1$. The main challenge arises from the stringent requirement of a good spatial overlap between a Rb BEC and a much larger K Fermi gas, as well as uncontrolled three-body losses when the Rb density gets too high. Furthermore, cooling KRb molecules directly in a harmonic trap is difficult due to its reactive nature.  Thus, it gradually became more attractive to create KRb molecules directly in a 3D optical lattice and optimize the filling fraction to realize a low entropy system.  A natural way to accomplish this task is to take advantage of the precise experimental control that is available for manipulating the initial atomic quantum gas mixture in the 3D lattice.  Specifically, we need to prepare low entropy states of both atomic species and and then utilize the already familiar techniques of coherent association and state transfer for efficient molecule production at individual lattice sites.  The combination of efficient magnetoassociation of preformed pairs~\cite{Chotia2012} and coherent optical state transfer via STIRAP means the second step should work well; however, creating a low entropy state for both species with the optimal density of one particle each per site is very challenging.  This idea was first proposed in 2003~\cite{Damski2003} and later specifically for the KRb system~\cite{Freericks2010}.  The creation of Rb$_2$ Feshbach molecules in a 3D lattice followed this basic idea, where the molecules were produced out of the region of a Mott insulator that has two atoms per site~\cite{Volz2006}. However, the KRb system faced a much bigger challenge due to the fact that we must address two different species with different mass and different quantum statistics.

The initial experimental target is to produce spatially overlapped atomic distributions that consist of a Mott insulator (MI) of Rb bosons~\cite{Greiner2002} and a spin-polarized band insulator of K fermions~\cite{Schneider2008}.  The occupancy of the MI depends on the ratio of the chemical potential $\mu$ to on-site interaction energy $U$, and can be higher than 1 per site if there are too many Rb atoms or if the underlying harmonic confinement of the lattice is too high.  The filling fraction of fermions in the optical lattice depends on an interplay between initial temperature, lattice tunneling, and external harmonic confinement. For our experiment, we need a sufficiently tight harmonic confinement and large K atom numbers to achieve a filling approaching unity in the center of the lattice.  The main challenge of the experiment is the opposite requirement between the filling of the Fermi gas and the number of Rb atoms that can be accommodated in the $n=1$ Mott insulator. This requires a careful compromise to achieve experimental optimization.

\begin{figure}
\begin{center}
\includegraphics[width=13cm]{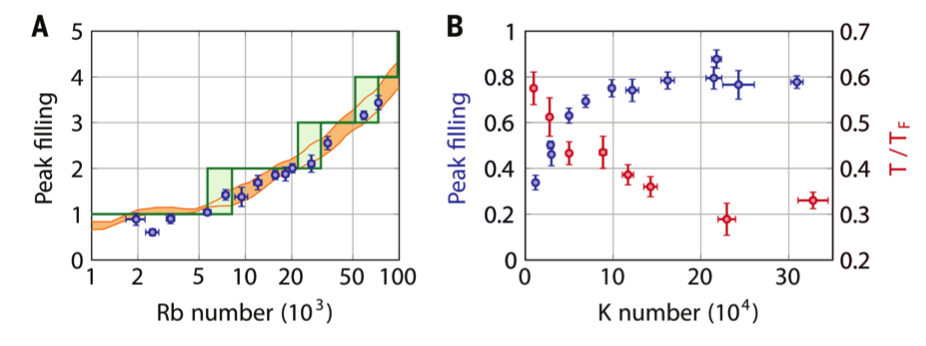}
\caption{(a) The peak filling of Rb as a function of the Rb number. The green lines show the zero temperature theory for filling, and the orange band shows the same theory predictions under a finite imaging resolution realized in the experiment.  (b) The peak filling in the lattice (blue circles, left vertical axis) and temperature of K prior to loading the lattice (red circles, right vertical axis) as function of K number. The temperature is normalized to the Fermi temperature of the K gas in the optical harmonic trap. Reproduced with permission from Reference~\cite{Moses2015}.}
\label{fig:fillingkandrb}
\end{center}
\end{figure}

\begin{figure}
\begin{center}
\includegraphics[width=13cm]{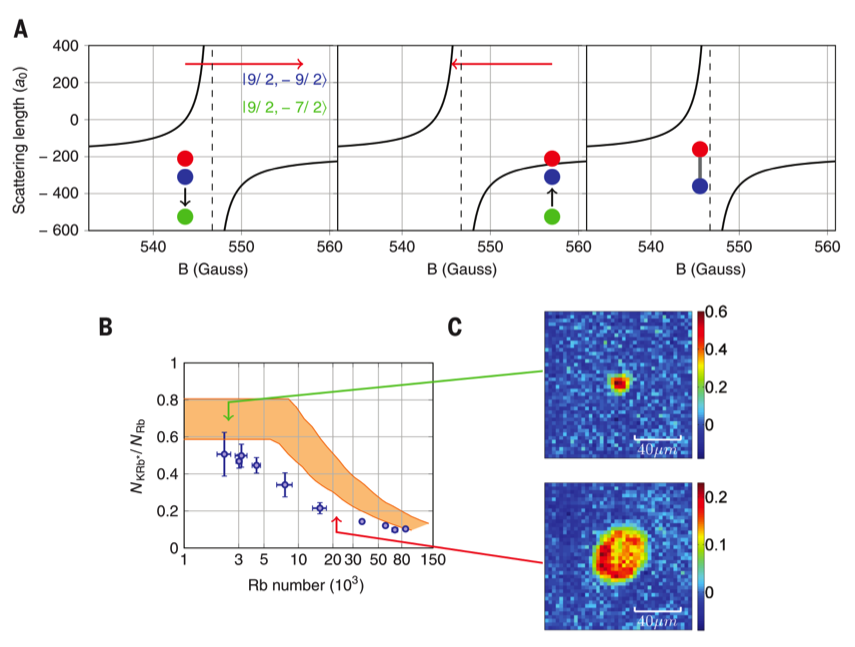}
\caption{(a) The K-Rb scattering length as a function of the magnetic field. Initially the K atoms are prepared in the resonant state $|9/2, -9/2 \rangle$.  We spin flip the K atoms to the non-resonant $|9/2, -7/2 \rangle$ state, before we sweep the magnetic field to above the Feshbach resonance, at which time we drive the K atoms back to the $|9/2, -9/2 \rangle$ state. We can then sweep the field to below the resonance to make Feshbach molecules.  (b) The number of Feshbach molecules produced normalized to the initial Rb number versus the number of Rb atoms. Theory band corresponds to calculations based on the zero temperature curve in Fig.~\ref{fig:fillingkandrb}a. (c) Images of clouds of ground-state molecules in the optical lattice for high and low filling fraction. Reproduced with permission from Reference~\cite{Moses2015}.}
\label{fig:fillingresult}
\end{center}
\end{figure}

In Reference~\cite{Moses2015}, we first studied the filling of the atomic gases separately and determined that we need to work with a very small MI (fewer than 5000 Rb atoms) and a large Fermi gas (more than $10^5$ K atoms) to achieve fillings approaching one atom of each species per site for the given external confinement potential (see Fig.~\ref{fig:fillingkandrb}). For the Fermi gas we reached the band insulator limit where the filling is saturated beyond $10^5$ K atoms.  In order to preserve the filling of the Rb MI in the presence of such a large K cloud, it is imperative to turn the interspecies interactions off by loading the lattice at $a_\text{K-Rb}=0$.  However, the magnetoassociation proceeds by ramping the magnetic field from high to low across the Feshbach resonance, while the location of $a_\text{K-Rb}=0$ is below the resonance. Hence, before we can perform magnetoassociation we need to first ramp the magnetic field from where $a_\text{K-Rb}=0$ is located to above the resonance.  To avoid populating higher bands when we do this, we first transferred K to the $\ket{9/2,-7/2}$ state to avoid the Feshbach resonance. The procedure is depicted in Fig.~\ref{fig:fillingresult}a. We then ramped the $B$ field above the resonance, and transferred the K atoms back to the $\ket{9/2,-9/2}$ state. Finally we converted the K-Rb pairs into Feshbach molecules on each site and we studied the fraction of Rb atoms that were converted to molecules (as Rb is the minority species). In the limit of a small Rb atom number, we converted more than 50\% of the Rb to molecules (Fig.~\ref{fig:fillingresult}b).  We then produced ground-state molecules and removed all unpaired atoms from the lattice. The final number of ground-state molecules can be determined by reversing the transfer process and then performing the normal absorption imaging on atoms.  Typical \textit{in situ} images are shown in Fig.~\ref{fig:fillingresult}c.  In the case of high conversion, we achieved molecule fillings of at least 25\%, which is significantly higher than our previous work~\cite{Yan2013, Hazzard2014}. With some quick experimental improvements we should be able to increase the filling even further.  This filling fraction is at the percolation threshold, where every molecule would be connected to every other molecule in the entire lattice~\cite{Moses2015}, and it marks the first time that the polar molecules have entered the quantum degenerate regime.  Studying the spin dynamics described in the previous section in this higher-filled lattice will be the subject of future work. We note that a similar procedure has recently been pursued in the RbCs experiment in Innsbruck, where a Rb superfluid was moved to overlap a Cs Mott insulator before increasing the lattice depth to produce a dual Mott insulator. This technique has lead to a gas of RbCs Feshbach molecules with 30$\%$ filling~\cite{Reichsollner2016}, but has not yet been combined with STIRAP to make ground-state molecules.  

One open question from this investigation is the efficiency with which a pair of K and Rb on a lattice site can be converted into a Feshbach molecule. Even if every site has one K and one Rb, a highly filled lattice of molecules can only be achieved if the conversion of free atoms to Feshbach molecules is very efficient. We already know that the STIRAP efficiency of converting Feshbach molecules to the ground state is about $90\%$. So we set out to develop another experimental technique to allow us to investigate this Feshbach molecule conversion efficiency.

\begin{figure}
\centering
\includegraphics[width=10cm]{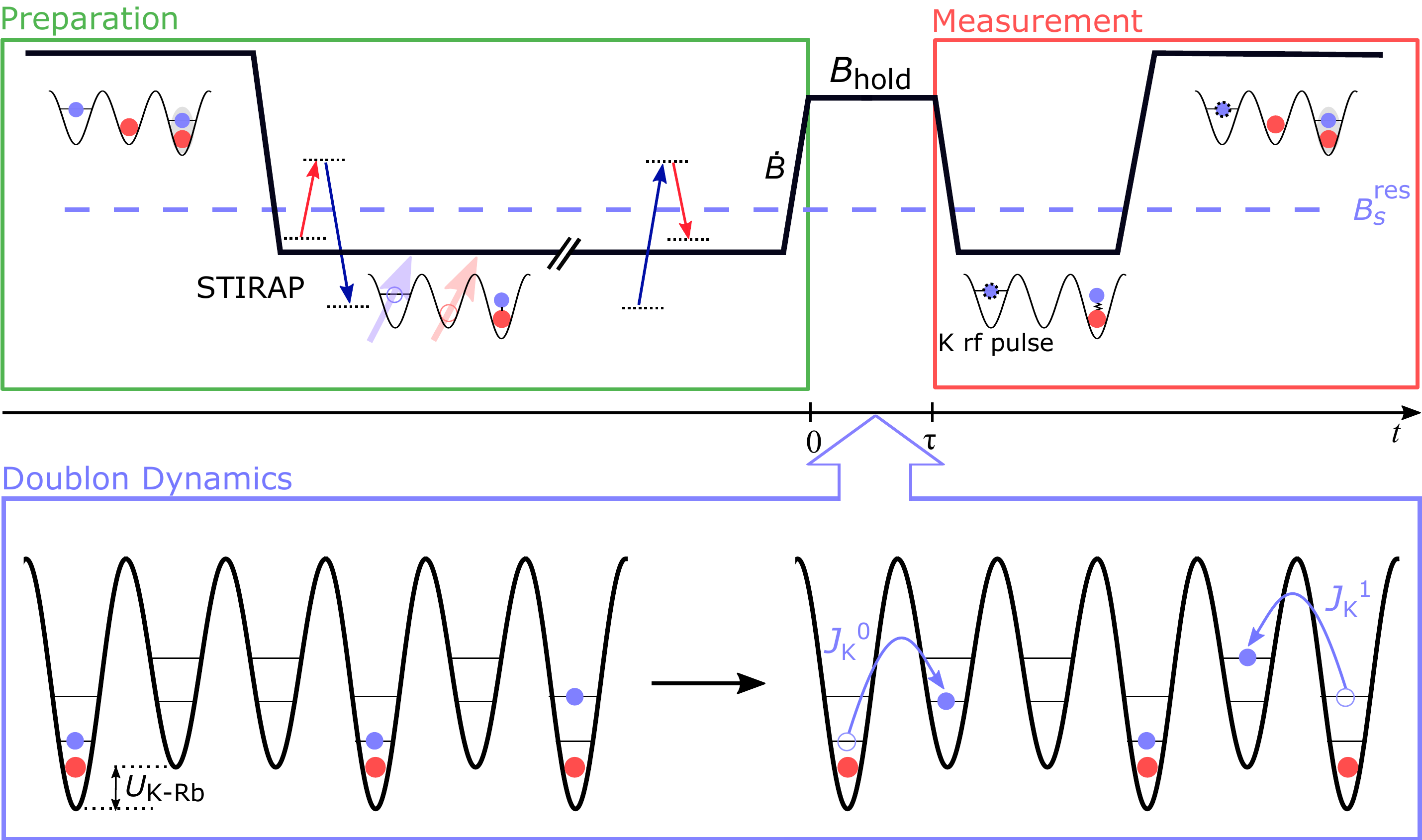}
\caption{The experimental setup for producing K-Rb doublons in an optical lattice. The first step is the preparation, where a pure sample of molecules in the optical lattice are turned into sites with one K and one Rb using the STIRAP sequence. We then let these doublons evolve in the lattice for a variable time at a variable scattering length. During the evolution, both K and Rb can tunnel (potentially together), but for short evolution times only K will tunnel since it feels a weaker lattice due its smaller mass. After the evolution period, we associate doublons that have not fallen apart back into Feshbach molecules, which can be spectroscopically differentiated in the measurement from unpaired K atoms that have separated from their doublon Rb partner. Reproduced with permission from Reference~\cite{Covey2016}.}
\label{Doublon}
\end{figure}

Once the KRb molecules are produced in the ground state, unpaired atoms can be removed with resonant light. In principle, this atom-removal process does not affect the molecules. However, the fact that we use a large number of K atoms and that KRb molecules can chemically react with K if they encounter each other at short range does lead to a loss of a small fraction of ground state molecules. After this process, the 3D optical lattice now has sites that are either empty or contain one ground-state molecule. The STIRAP process can then be reversed, which leads to sites that are either empty or contain a Fesbach molecule. Then, we can sweep the field back above the Feshbach resonance to dissociate the molecules, to create a scenario where lattice sites are either empty or contain a Bose-Fermi atomic pair, which we refer to as a `doublon'. This process is outlined in Fig.~\ref{Doublon}, and the preparation of this clean initial condition can be used to study the conversion efficiency of doublons into Feshbach molecules.

We have identified effects from a narrow, higher order $d$-wave Feshbach resonance near the broad $s$-wave resonance we use for molecule association. This narrow resonance was causing our Feshbach molecule creation efficiency to be $70\%$ since we first produced molecules in an optical lattice~\cite{Chotia2012}. By simply ramping across the resonance sufficiently fast so that we avoid making $d$-wave molecules adiabatically, we are able to solve this problem. With this mechanism understood, we anticipate $100\%$ conversion efficiency for all future experiments. We want to emphasize that since the $d$-wave resonance is only 9 mG wide, it would play no role in molecule formation in a harmonic dipole trap. However, since the on-site density in an optical lattice is so high, the timescales for keeping the sweep diabatic with respect to this resonance are much faster. We believe this to be the only investigation to date of narrow heteronuclear Feshbach resonances in an optical lattice. We note, however, that similar resonances have been explored in Cs$_2$, and navigation through the state manifold was carried out in Ref.~\cite{Danzl2010}. 

We have also investigated tunneling dynamics and interaction effects between the bosonic and fermionic atoms composing the doublons. Moreover, if sufficient tunneling time is allowed, we can even observe effects of doublon-doublon interactions. This type of system could be very interesting for the study of quasi-crystallization and many-body localization in an optical lattice (see Ref.~\cite{Covey2016}).

\section{Outlook: new directions}
With strong scientific impact to several different disciplines, the future prospect for the field of ultracold polar molecules is bright. We will present several exciting directions in which current research is now heading. The first immediate opportunity is to further increase the phase-space density of the molecular system and thus reduce the entropy. Such improvements are essential to the realization of certain spin models that are currently of great interest in the condensed-matter physics community. The second direction addresses new tools for manipulating and detecting polar molecules with higher precision, which will allow for a new generation of experiments for both quantum physics and chemistry. The third direction is to create more complicated and interesting molecules with an even richer internal structure and more complex interactions.

\subsection{Evaporative cooling of reactive dipolar molecules in quasi-2D}
One possible way to increase the phase-space density of polar molecular gases is to perform evaporative cooling, as is commonly done for many atomic species~\cite{Anderson1995, Davis1995, Bradley1997}. The scattering and collisions of molecules are obviously much more complicated than atoms, and since efficient evaporative cooling relies entirely on thermalization via collisions, it is not clear that it would work well for molecules. Indeed, the molecular collisions must be well understood in order to construct an efficient evaporation trajectory~\cite{Zhu2013}. In a magnetically trapped gas of the hydroxyl radical, specific quantum states can be used to exhibit elastic collisions mediated by van der Waals interactions~\cite{Stuhl2012}.

The system of ultracold polar fermionic molecules offers the additional complexity that $s$-wave collisions are not allowed, and $p$-wave collisions are suppressed at ultralow temperatures due to the centrifugal barrier. This makes efficient thermalization really challenging even if the molecules are not reactive. However, as was discussed in Sec.(2.4), polar molecules can be confined in a 2D spatial geometry with an external electric field applied perpendicular to the 2D plane to significantly suppress undesired lossy collisions. At the same time, electric-field induced dipole-dipole interactions mix even and odd collision partial waves. Thus, the long-range interactions of molecules can be expanded in the canonical angular momentum scattering basis as a linear combination of partial waves, and as a result even identical fermions can have significant scattering amplitudes at ultralow collision energies.

This is referred to as universal dipolar scattering, and it has been realized in ultracold atom experiments with the highly magnetic lanthanide atoms that interact with magnetic dipoles. Such scattering has lead to the direct cooling of spin-polarized degenerate Fermi gases of Dy~\cite{Lu2012} and Er~\cite{Aikawa2014}, without the need of having other species to sympathetically cool the fermions at ultralow temperatures.

\subsection{Putting molecules under the microscope (high resolution imaging)}
\begin{figure}
\centering
\includegraphics[width=3in]{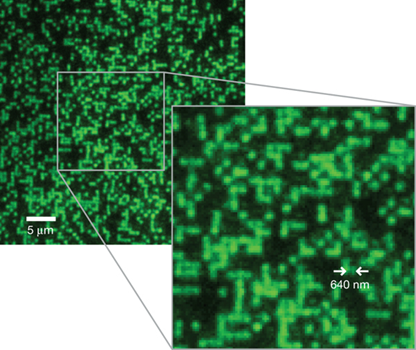}
\caption{Images of single atoms in a Rb Bose-Einstein condensate trapped in an optical lattice. The imaging resolution is sufficient to resolve atoms on neighboring lattice sites since the point-spread function of individual atoms is smaller than the lattice spacing. Reproduced with permission from Reference~\cite{Bakr2009}.}
\label{microscope}
\end{figure}

The technology of working with ultracold atoms in optical lattices has matured greatly over the past decade, with the precise control of atomic interactions and various configurations of the lattice geometry and tunneling energy. These capabilities have become very useful when lattice-based cold atom systems are being developed to realize novel Hamiltonians that are interesting of their own merit or useful to simulate some outstanding problems in condensed matter phenomena~\cite{Greiner2008, Bloch2012}. The dipole-dipole interactions present in ultracold polar molecules offer exciting extensions to these models in which long-range interactions play a central role~\cite{Carr2009, Gorshkov2011b}.  Already, an extended Bose-Hubbard model, where dipole-dipole interactions modify the phase diagram, has been studied using magnetic atoms~\cite{Baier2016}. An exciting new technology with ultracold atoms in optical lattices is the development of high resolution spatial detection of the atoms in the lattice~\cite{Gemelke2009}, where several groups are now able to resolve and image atoms located on individual lattice sites~\cite{Bakr2009, Sherson2010, Omran2015, Parsons2015, Cheuk2015, Haller2015}, as shown in Fig.~\ref{microscope}. Such techniques are referred to as quantum gas microscopy.

It will be tremendously powerful if we can extend the quantum gas microscope tool to ultracold polar molecules so that we have single-site imaging resolution for molecules in an optical lattice. In this particular area, coherent association of cold atoms to form ultracold molecules offers an enormous advantage over direct cooling of molecules. Imaging of ultracold polar molecules can be easily accomplished by dissociating the molecules into their constituent atoms, which can be efficiently detected by scattering thousands of photons. Therefore, a quantum gas microscope for polar molecules would be operated the same way as for atoms, except that the molecules are dissociated prior to imaging.

High resolution imaging of polar molecules in an optical lattice will pave the way for investigations of non-equilibrium many-body dynamics, such as spin diffusion and many-body localization, where individual excitations can be tracked as they propagate throughout the system. Moreover, this technique can be extended to address localized regions of the lattice to create an array of out-of-equilibrium molecules with excitations~\cite{Weitenberg2011}.

\subsection{Large, stable electric fields provided with in-vacuum electrodes}
\begin{figure}
\centering
\includegraphics[width=3in]{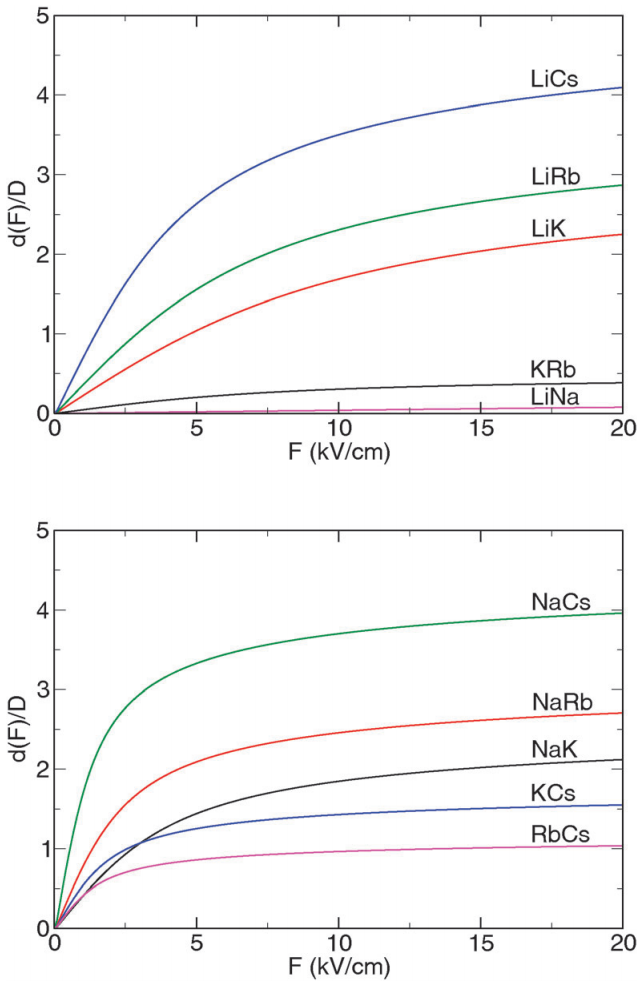}
\caption{The dipole moments for all experimentally relavent bialkali ground-state molecules. The top panel includes molecules that are chemically unstable with respect to two-body exchange reactions, while the bottom shows molecules that are expected to be stable against such reactions. Reproduced with permission from Reference~\cite{Julienne2011}.}
\label{dipolemoments}
\end{figure}

For most polar molecules, an electric field of several kV/cm is required to induce a significant dipole moment. References~\cite{Julienne2011, Quemener2012} have a table of all the bialkali combinations which shows that most species are not fully polarized until 5 kV/cm, and some (like KRb) are not fully polarized until $>40$ kV/cm (see Fig.~\ref{dipolemoments}). One simple reason for this requirement of a large field is the lack of parity-doublets in the singlet ground-state potential for bialkalis, where only the opposite parity rotational states can be used for molecular polarization. The energy spacing for neighboring rotational states is typically a few GHz. Therefore, the design of electrodes for ultracold polar molecules experiments must be done carefully with the specific molecule of choice in mind.

Once polarized, polar molecules become very sensitive to inhomogeneities and instabilities of the electric field. This sensitivity can be a challenge for experiments, and the control of the field, in both the spatial and time domains, becomes exceedingly important if precise manipulations of molecules using electric field are a desired feature. Much of our early work with ultracold polar molecules in electric fields was done with electrodes located outside the vacuum chamber. Indeed, we had to deal with extensive complications due to transient electric fields, which were attributed to charge build-up and polarization of the vacuum glass cell.

An obvious solution is to set up high voltage electrodes inside the vacuum system, and indeed a number of research groups have started along this path (there are few publications at the time of this writing, but see for example~\cite{Grobner2016}). Typical configurations are composed of metal rods and/or glass plates coated with a conductive thin film like indium tin oxide (ITO). Such semiconductor thin films are known to be damaged by alkali vapor~\cite{Daschner2012}, and so care must be taken to limit the alkali flux onto the coating to a sufficiently low level. Alternatively, metal thin films like nickel work well with alkalis, but the optical transmission is significantly lower for the same thickness. Metal plates would be more convenient, but they prohibitively limit the available optical access. Metal rods are typically arranged in a quadrupole fashion, which is a versatile configuration for the application of fields with gradients or somewhat homogeneous fields.

Naturally, the field from rods alone will not be perfectly homogeneous in the space surrounded by the rods, and one must ensure that the field inhomogeneity experienced by the entire molecular cloud does not present a problem for the experiment of interest. Since some of our next experimental goals are to investigate the full spin-1/2 Hamiltonian that incorporates both the Ising and spin-exchange interactions~\cite{Yan2013}, we must ensure that the site-to-site Stark shift for molecules in the optical lattice is significantly smaller than the molecule interaction energy. We have therefore decided to pursue a combination of ITO-coated plates and four metal rods in a configuration that will allow a homogeneous field to be present in the central region while still being able to accommodate the application of electric field gradients when necessary. The electric field gradient can be used to spectroscopically select particular lattice sites for example, or be used to tilt the 2D optical trap to enable efficient evaporation of molecules.

\subsection{Detection of cold chemical reaction products}
So far, chemical reactions in the quantum regime have been probed only through the loss measurement of the reactants and the dependences on certain quantum properties for the entrance channel~\cite{Ospelkaus2010b,Ni2010,Miranda2011}; the final products of such reactions have not been directly observed. Quantum chemists are very interested in such investigations since they would shed light on the specifics of the energy and momentum transfer between the reactants and products in a fully quantized system~\cite{Nesbitt2012}.  Detection of the products in this type of quantum gas experiments is very challenging for many reasons. First, the products typically have significantly more kinetic energy than the depth of the trap that is used to confine the reactant molecules, and so the product molecules are lost too quickly to detect. Second, the product molecules cannot be probed by scattering many photons as in a typical atomic absorption or fluorescence imaging. Third, the goal is to identify the specific quantum state distributions of the products, and therefore any detection technique must be state specific. The best way to address such requirements and limitations is to state-selectively photo-ionize the product molecules, which can then be detected with ion detectors inside the vacuum chamber. No experiments with ultracold polar molecules have ion detectors to date, and so such experiments must specifically be designed with these goals. Efforts are underway in the group of Kang-Kuen Ni at Harvard University to detect the quantum states of the products from reactions of ultracold KRb molecules.

\subsection{Molecules with both electric and magnetic dipoles}
Many research groups are now pursuing ultracold molecules composed of more exotic atoms, including alkaline earth and rare earths. Combinations of alkalis with alkaline earths produce diatomic radicals, which have an unpaired electron. Thus, such heteronuclear molecular species have both an electric dipole moment and a magnetic dipole moment, and offer researchers even more possibilities for control~\cite{Stuhl2012, Stuhl2013}. Notable examples of recent work with ultracold mixtures include efforts towards RbSr~\cite{Pasquiou2013}, DyK, DyEr (both underway at the University of Innsbruck), and LiYb~\cite{Dowd2015} molecules. Most experiments designed to cool and trap molecules either with a Stark decelerator or a magneto-optical trap use molecules with a single unpaired electron, such that the interactions with light fields nearly mirror that of alkali atoms~\cite{Stuhl2008, Shuman2009}. Examples of such molecules include SrF~\cite{Barry2014}, OH~\cite{Stuhl2012, Stuhl2013}, YO~\cite{Hummon2013, Yeo2015}, YbF~\cite{Hudson2011}, ThO~\cite{Baron2014}, CaF~\cite{Tarbutt2015} and many others.

\bibliographystyle{unsrt}

\end{document}